\documentclass[aip,jcp,amsmath,amssymb,reprint,floatfix]{revtex4-1}
\usepackage{graphicx}
\usepackage[capitalize]{cleveref}
\crefname{fig}{Fig.}{Figs.}
\Crefname{fig}{Figure}{Figures}
\crefname{tab}{Table}{Tables}
\Crefname{tab}{Table}{Tables}
\crefname{alg}{Algorithm}{Algorithms}
\Crefname{alg}{Algorithm}{Algorithms}
\crefname{sec}{Sec.}{Secs.}
\Crefname{sec}{Section}{Sections}
\crefname{eq}{Eq.}{Eqs.}
\Crefname{eq}{Equation}{Equations}

\begin{document}

\title{Control of probability flow in Markov chain Monte Carlo---Nonreversibility and lifting}

\author{Hidemaro Suwa}
\email[]{suwamaro@phys.s.u-tokyo.ac.jp}
\affiliation{Department of Physics, The University of Tokyo, Tokyo 113-0033, Japan}

\author{Synge Todo}
\email[]{wistaria@phys.s.u-tokyo.ac.jp}
\affiliation{Department of Physics, The University of Tokyo, Tokyo 113-0033, Japan}
\affiliation{Institute for Physics of Intelligence, The University of Tokyo, Tokyo 113-0033, Japan}
\affiliation{Institute for Solid State Physics, The University of Tokyo, Kashiwa 277-8581, Japan}

\date{\today}

\begin{abstract}
The Markov chain Monte Carlo (MCMC) method is widely used in various fields as a powerful numerical integration technique for systems with many degrees of freedom.
In MCMC methods, probabilistic state transitions can be considered as a random walk in state space, and random walks allow for sampling from complex distributions.
However, paradoxically, it is necessary to carefully suppress the randomness of the random walk to improve computational efficiency.
By breaking detailed balance, we can create a probability flow in the state space and perform more efficient sampling along this flow.
Motivated by this idea, practical and efficient nonreversible MCMC methods have been developed over the past ten years.
In particular, the lifting technique, which introduces probability flows in an extended state space, has been applied to various systems and has proven more efficient than conventional reversible updates.
We review and discuss several practical approaches to implementing nonreversible MCMC methods, including the shift method in the cumulative distribution and the directed-worm algorithm.
\end{abstract}


\maketitle

\section{Introduction}
\label{sec:introduction}

The dimensionality of state variables increases proportionately to the number of particles or sites.
The Markov chain Monte Carlo (MCMC) method is particularly powerful for high-dimensional problems and is essential for studying phase transitions, critical phenomena, and dynamics in interacting systems.\cite{newman_monte_1999,landau_guide_2014}
In MCMC, states are updated using transition probabilities that depend on the current state, generating a Markov chain.
As a result, a long-time average enables the sampling of state variables from arbitrary distributions (target distributions).
Such state transitions can be regarded as a kind of random walk in state space.
States with higher weights (measures) appear more frequently in the Markov chain.

Although random walks enable state generation from an arbitrary distribution, the randomness of sampling lowers the computational efficiency.
Paradoxically, reducing the randomness of the random walk helps improve efficiency.
This involves creating a probability flow in state space and sampling efficiently along this flow.
A helpful analogy is mixing milk in coffee: It mixes faster when stirred than when left to diffuse.
To quickly mix biased initial states, it is necessary to optimize the overall probability flow while maintaining balance.
Motivated by this mechanism, recent efforts have been made to introduce and optimize probability flows.
The existence of a probability flow in state space breaks reversibility and produces a nonreversible Markov chain.
Probability flows significantly improve error scaling with respect to the number of particles or sites in several systems, such as the one-dimensional Ising~\cite{sakai_dynamics_2013} and Potts~\cite{Faizi2020} models and the three-dimensional Heisenberg model.~\cite{Nishikawa2015}
There is no mathematical theorem that excludes the improvement of the scaling exponent in the other systems.
Even when the power of the error scaling is the same as that of the corresponding reversible chain, various efforts to control probability flows can significantly reduce the prefactor of the scaling in many cases.
We review and discuss approaches that introduce and control probability flows that enable efficient sampling in the nonreversible MCMC method.

\section{Balance condition}

One of the fundamental algorithms in the MCMC method is the Metropolis algorithm proposed by Metropolis et al.\ in 1953 (see Ref.~\onlinecite{metropolis_equation_1953}).
Many MCMC algorithms, including the Metropolis and the heat bath algorithms (Gibbs sampler),\cite{creutz_monte_1980,geman_stochastic_1984} satisfy detailed balance, namely reversibility:
\begin{equation}
\pi_i P_{ij} = \pi_j P_{ji} \quad \forall i, j,
\end{equation}
where $\pi_i$ is the weight of state $i$ (the measure of the target distribution) and $P_{ij}$ is the transition probability from state $i$ to state $j$.
These reversible MCMC methods simulate physical dynamics that seem plausible, as real physical systems also satisfy detailed balance in equilibrium.
However, if the goal is numerical integration using running averages, Monte Carlo dynamics are not constrained by physical requirements.
Detailed balance is not necessary, and global balance, which ensures that the target distribution is stationary under time evolution, \cite{robert_monte_2004} suffices\footnote{In the present paper, we assume irreducibility of the Markov chain, which ensures, under global balance, convergence of the running average to the expectation value with respect to the target.}
\begin{equation}
\sum_j \pi_i P_{ij} = \sum_j \pi_j P_{ji} \quad \forall i.
\label{eq:gb0}
\end{equation}
This condition means the total probability flow into and out of each state balances.
Detailed balance is a sufficient but not necessary condition for this global balance.
MCMC methods that satisfy global but not detailed balance are called nonreversible MCMC methods.
Physically, this corresponds to calculating long-time averages using nonequilibrium steady states.

Interestingly, it has been proven that breaking the reversibility in a reversible method always speeds up distribution convergence.\cite{hwang_accelerating_2005,ichiki_violation_2013,duncan_variance_2016}
The problem of how to optimally introduce probability flow remains a hot topic in probability theory and applied mathematics.
In the following, we discuss probability optimization and lifting, relatively simple methods to construct efficient nonreversible MCMC methods.

\section{Computational efficiency of the MCMC method}
\label{sec:efficiency}

In the MCMC method, a generated state depends on the previous states, leading to correlations (autocorrelations) between the states at different times.
Generally, reducing autocorrelation improves the computational efficiency of the MCMC method.
The autocorrelation function, defined as
\begin{equation}
A_\mathcal{O}(t) = \frac{\langle \mathcal{O}_i \mathcal{O}_{i+t} \rangle - \langle \mathcal{O}_i \rangle^2}{\langle \mathcal{O}_i^2 \rangle - \langle \mathcal{O}_i \rangle^2},
\label{eq:A}
\end{equation}
describes the correlation of physical quantities at time $t$ in Monte Carlo steps.
Here, $\mathcal{O}_i$ represents the value of a physical quantity, or observable, $\mathcal{O}$ at the $i$th Monte Carlo step. 
The bracket $\langle \cdot \rangle$ denotes the Monte Carlo average, taking the average over $i$, and thus the autocorrelation function~\eqref{eq:A} is independent of $i$.
In most cases, the autocorrelation function is given by the sum of multiple exponential functions,~\cite{Sokal1997} and at large $t$, it decays as
\begin{equation}
A_\mathcal{O}(t) \approx e^{-t / \tau_{\text{exp}, \mathcal{O}}},
\end{equation}
where $\tau_{\text{exp}, \mathcal{O}}$ is the exponential autocorrelation time or relaxation time, indicating the degree of correlation with previous values.
On the other hand, the mean squared error of physical quantities obtained from the MCMC method asymptotically follows
\begin{equation}
\sigma_\mathcal{O}^2 \approx \frac{v_{\rm asymp,\mathcal{O}}}{M},
\label{eq:variance}
\end{equation}
where $M$ is the total number of samples, and the asymptotic variance $v_{\text{asymp}, \mathcal{O}}$ quantifies the sampling efficiency, being often used for performance comparisons of algorithms.
Assuming no bias in physical quantity calculations, the asymptotic variance can be expressed as
\begin{equation}
  v_{\text{asymp}, \mathcal{O}} = 2 \tau_{\text{int}, \mathcal{O}} v_\mathcal{O},
\label{eq:variance2}
\end{equation}
where $v_\mathcal{O}$ is the variance of the physical quantity, and the integrated autocorrelation time is defined as
\begin{equation}
  \tau_{\text{int}, \mathcal{O}} = \frac{1}{2} + \sum_{t=1}^{\infty} A_\mathcal{O}(t).
\end{equation}
From \cref{eq:variance,eq:variance2}, it is clear that the effective number of samples decreases from $M$ to $M / 2\tau_{\text{int}, \mathcal{O}}$ due to autocorrelation.
When the autocorrelation function is a single exponential and the correlation times are large, the integrated autocorrelation time is almost identical to the exponential autocorrelation time.

On the other hand, at the critical point, critical slowing down causes the autocorrelation time to increase proportionally to $L^z$, where $L$ is the system size, and $z$ is the dynamic critical exponent.
In local reversible update methods, such as single-particle or single-site updates using the Metropolis algorithm, $z$ typically becomes around 2.
Although critical slowing down is a physical phenomenon, the long autocorrelation time becomes a barrier to phase transition analysis.
Reducing the dynamic critical exponent is crucial for analyzing critical phenomena.

These autocorrelation times and the resulting dynamic critical exponents depend on the observables to measure. 
However, the exponential autocorrelation time for most observables is given by $\tau_{{\rm exp}, \mathcal{O}}=-1/\ln | \lambda_2|$, where $\lambda_2$ is the second largest eigenvalue of the transition matrix, independent of the observable.
Furthermore, the second largest eigenvalue is associated with the mixing time, in which any initial distribution nearly converges to the target $\pi$ within a small tolerance in the total variation distance.
The mixing time is typically $\tau_{\rm mix} \sim -1/\ln | \lambda_2|$, giving an estimate of an appropriate period of thermalization, or burn-in, steps.
In practice, the integrated autocorrelation time is relatively easy to estimate, and we can approximately estimate the exponential autocorrelation time and an appropriate period of thermalization steps.

\section{Shift in cumulative distribution}
\label{sec:ags}

For discrete state space, several approaches to introducing probability flow have been proposed, and their effectiveness has been tested in representative models, such as the Ising and Potts models.
In particular, the shift in the cumulative distribution naturally introduces a single parameter, the amount of the shift, and provides a handy approach to controlling the rejection rate and probability flow.\cite{suwa_markov_2010,todo_geometric_2013,suwa_reducing_2024}
Here, we review the shift approach and discuss an extension to continuous variables.

\begin{figure}
  \begin{center}
\includegraphics[bb=0 0 1418 1292,width=0.7\columnwidth]{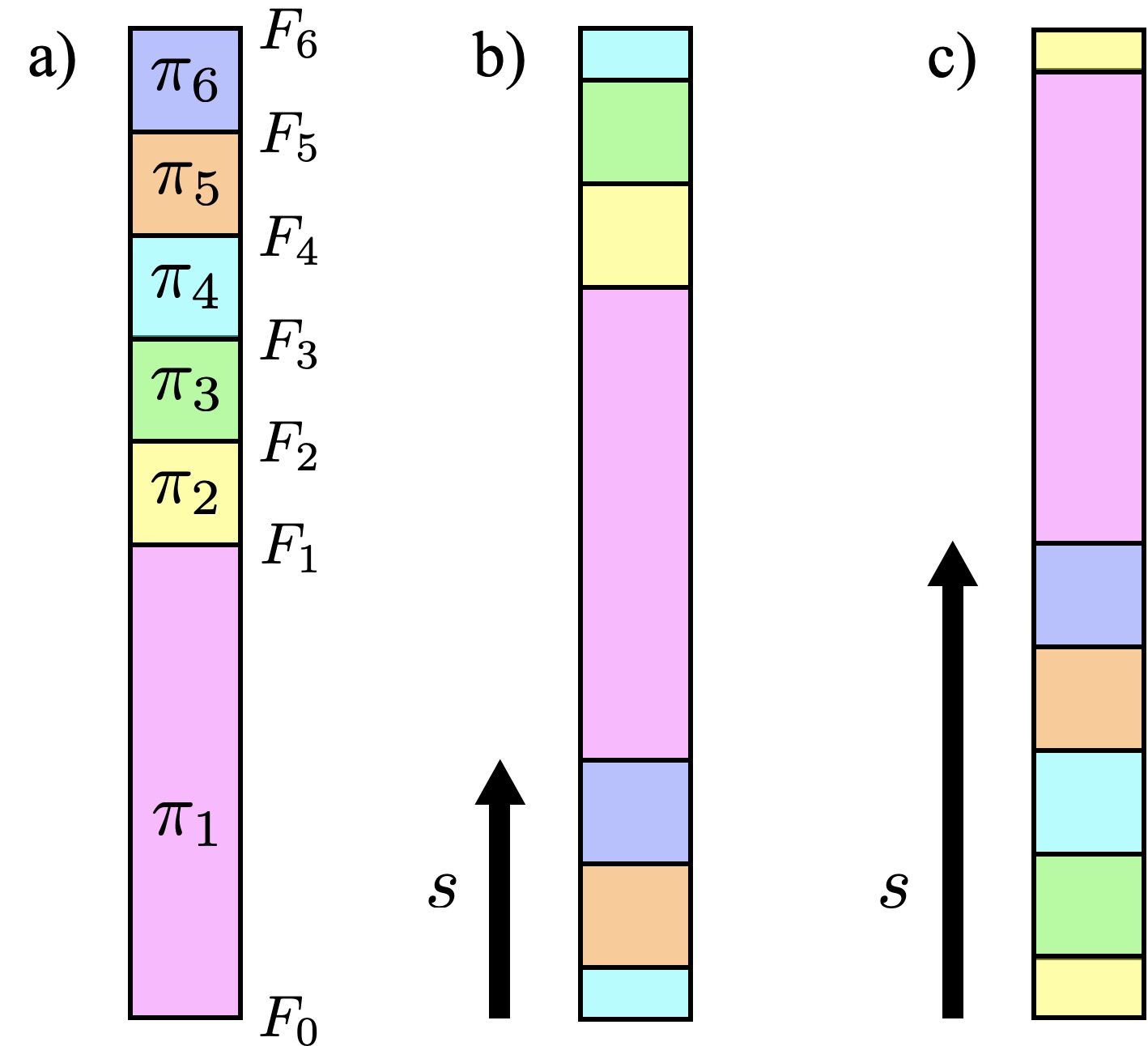}
\caption{\label{fig:shift}(a) Example of the weight tower for $n=6$ and (b) a periodic shift of the tower. 
$\pi_i$ is the weight of state $i$ and $F_i$ is the cumulative distribution.
The parameter $s$ represents the shift amount of the weight tower.
The stochastic flows and the transition probabilities are determined from the overlap between the original and shifted towers for each state. 
See the main text for the details.
(c) When $s=\max_i \pi_i$, this approach reduces to the Suwa--Todo algorithm.~\cite{suwa_markov_2010}
This figure was taken from Ref.~\onlinecite{suwa_reducing_2024}.
}
  \end{center}
\end{figure}

Let us consider an update of a state that can take one of $n$ states.
For example, in the single-site update of the six-state Potts model, $n=6$.
We first calculate the cumulative weight distribution,
\begin{equation}
 F_i= \sum_{j=1}^i \pi_j \qquad (1 \leq i \leq n)
\end{equation}
and $F_0=0$. 
Here, the order of the states is arbitrary. 
Calculating the cumulative distribution is nontrivial for a huge $n$ but easy for a small $n$, such as in the update of local variables.
Let us next consider the sequence of $F_i$, or the weight tower, as shown in Fig.~\ref{fig:shift}~(a) for $n=6$.
We then perform shifting the tower by a certain amount, as shown in Fig.~\ref{fig:shift}~(b).
The shift is periodic; the weight shifted over the top of the tower is allocated to the lowest part of the tower.

Using this weight shift, we determine the transition probability from the overlap between the original and shifted towers.
It is convenient to define the stochastic flow from $i$ to $j$: $v_{ij}= \pi_i P_{ij}$.
Let $s$ denote the amount of the shift, $0 < s < F_n$.
We focus on the overlap between the ranges $[F_{j-1}, F_j]$ and $[F_{i-1} + s, F_i + s]$. 
The stochastic flows $v_{ij}$ are set to the overlap for $i=1,2$ and $3$ in the case of Fig.~\ref{fig:shift}(b).
If $F_i + s > F_n$, we take into account an additional overlap between the ranges $[F_{j-1}+F_n, F_j+F_n]$ and $[F_{i-1} + s, F_i + s]$, assuming that the tower is periodic, for example, for $i=4,5$, and $6$, as shown in Fig.~\ref{fig:shift}(b).
If a shifted weight range crosses the top of the original tower, nonzero flows are allocated from the overlaps in the above-mentioned two cases, such as for $i=4$ shown in Fig.~\ref{fig:shift}(b) and $i=2$ shown in Fig.~\ref{fig:shift}(c).
The resulting stochastic flow set by the periodic shift $s$ becomes\cite{suwa_reducing_2024}
\begin{eqnarray}
  v_{ij} \!\! &=& \!\! \max(0, \ \min( \Delta_{ij}, \ \pi_i + \pi_j - \Delta_{ij}, \ \pi_i, \ \pi_j)) \label{v_ij} \\
  && \hspace{-4.5mm} + \max(0, \ \min( \Delta_{ij} - F_n, \ \pi_i + \pi_j + F_n - \Delta_{ij}, \ \pi_i, \ \pi_j)) \notag,
\end{eqnarray}
where
\begin{equation}
\Delta_{ij} = F_i - F_{j-1} + s. \label{delta}
\end{equation}
The first term of Eq.~\eqref{v_ij} is the overlap between the ranges $[F_{j-1}, F_j]$ and $[F_{i-1} + s, F_i + s]$, and the second term is the overlap between the ranges $[F_{j-1}+F_n, F_j+F_n]$ and $[F_{i-1} + s, F_i + s]$.
Reference~\onlinecite{suwa_reducing_2024} should be referred for the derivation of the analytical expression.
The transition probability is then set to $P_{ij}=v_{ij}/\pi_i$.

Clearly, each range of $\pi_i$ in the original tower is covered by the shifted tower due to the periodic shift.
This shift algorithm always satisfies the following condition:
\begin{equation}
    \pi_i = \sum_j v_{ji} \qquad \forall i \label{eq:gb2},
\end{equation}
equivalent to the global balance~\eqref{eq:gb0}.
Using the shift algorithm for each local variable satisfies the global balance of the total system.~\cite{suwa_reducing_2024}

The rejection-free condition is straightforward in the shift approach: $v_{ii}=\pi_i P_{ii}=0$ $\forall i$, can be obtained for $\pi_{\rm max} \leq s \leq F_n - \pi_{\rm max}$ if $\pi_{\rm max} \leq F_n/2$, and there is no rejection-free solution otherwise.
When $s=\max_i \pi_i$, as shown in Fig.~\ref{fig:shift}(c), this approach reduces to the Suwa--Todo algorithm.~\cite{suwa_markov_2010}
Setting $s=F_n/2$ is another optimal choice, minimizing the rejection probability.
It provides one of the best local update methods in the Potts model.
Thanks to the introduction of the shift parameter, we can readily control the rejection probability.
The autocorrelation time of the order parameter in the Potts model exponentially decreases with the reduction of the rejection rate.~\cite{suwa_reducing_2024}

Next, let us consider an update of a continuous state variable for which the inversion method can be used.
It can be updated by using the heat bath algorithm, namely, the Gibbs sampler; a uniformly random variable $r \in [0,1]$ is generated, and the next state is chosen from the inverse function of a conditional cumulative distribution.
We can extend the shift approach to this case.
In order to explain the shift in a continuous case, let us consider the bivariate Gaussian distribution as a simple example,
\begin{equation}
  \pi(x_1, x_2) \propto e^{- \frac{(x_1-x_2)^2}{2\sigma_1^2}- \frac{(x_1+x_2)^2}{2\sigma_2^2}}.
\end{equation}
Given $x_2$, the local variable $x_1$ is updated by using the
conditional (cumulative) distribution,
\begin{equation}
  F(x_1 | x_2 ) = \int_{-\infty}^{x_1} \pi(x, x_2)dx.
\end{equation}
The Gibbs sampler determines the next state as
\begin{equation}
x'_1 = F^{-1}( r ),
\end{equation}
where $r\in [0,1]$ is a uniformly (pseudo)-random variable.
This process satisfies the detailed balance.

The over-relaxation method~\cite{adler_over-relaxation_1981} is known to be one of the best ways to update Gaussian variables.
The name over-relaxation comes from the idea of making the Markov chain have a negative correlation.
In this method, for the generation of a variable from a conditional Gaussian distribution $ \pi(z_i | \, \cdot \,) \sim \mathcal{N}(\mu_i, \sigma_i^2)$, the next state is chosen as $ z'_i = \mu_i + \alpha( z_i - \mu_i ) + \sigma_i \sqrt{ 1 - \alpha^2} \nu$, where $\nu$ is a random variable generated from $\mathcal{N}(0,1)$ and $\alpha$ is a parameter ($-1 < \alpha < 1$).
For more than two Gaussian variables, each variable can be updated sequentially using the conditional distribution, similar to the bivariate Gaussian variables.
Extending the over-relaxation method to a general distribution is an interesting problem.
One of the extended approaches, called ordered over-relaxation, was proposed;\cite{neal_suppressing_1998} after some candidates are generated and ordered, the next state is chosen on the approximately opposite side from the current position.

\begin{figure}[tbp]
  \begin{center}
  \includegraphics[width=0.95\columnwidth]{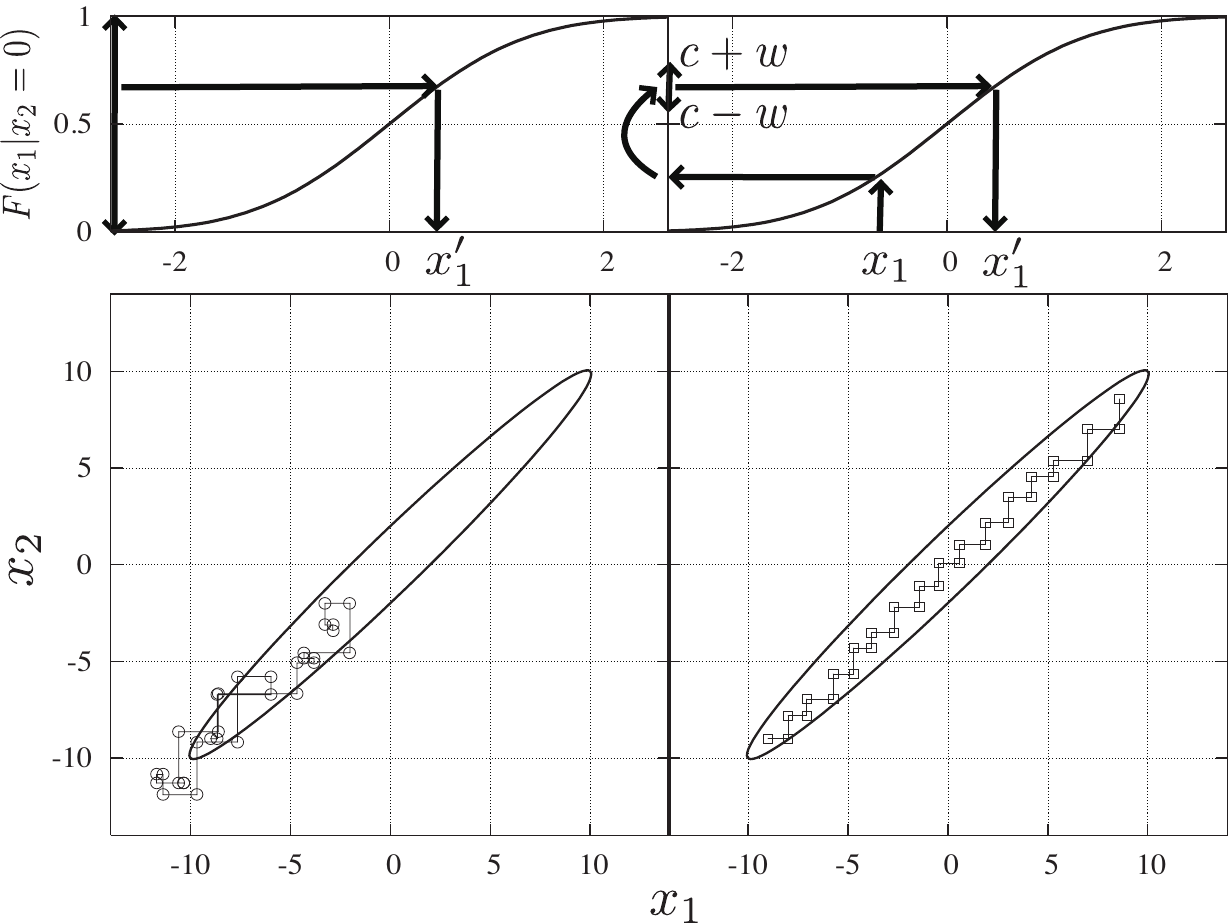}
  \caption{Trajectories of configurations updated by the Gibbs sampler (left) and by the present nonreversible algorithm with $c=0.4$ and $w=0.1$ (right) in the bivariate Gaussian distribution with $\sigma_1=1$ and $\sigma_2=10$.
  The ellipsoidal line is the three-sigma line of the Gaussian distribution.
  The upper figures show the update procedures of each algorithm.
  This figure was taken from Ref.~\onlinecite{suwa_geometrically_2014}.}
  \label{fig:gaussian-shift}
  \end{center}
\end{figure}

\begin{figure}[tbp]
  \begin{center}
  \includegraphics[width=1.0\columnwidth]{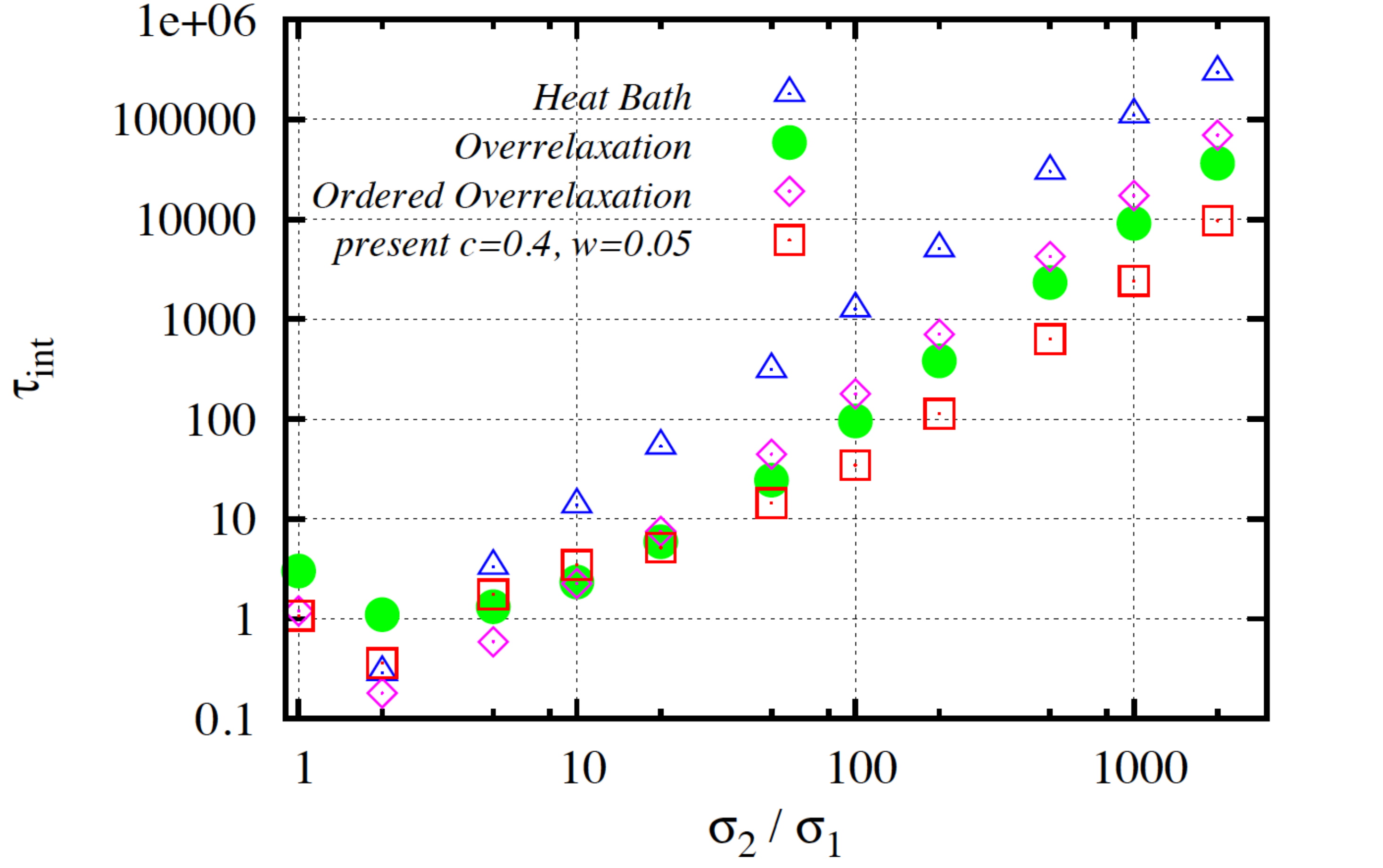}
  \caption{Autocorrelation times of $(x_1 + x_2 )^2$ in the bivariate Gaussian distribution by using the Gibbs sampler (triangles), the over-relaxation (circles) with $\alpha=-0.86$, the ordered over-relaxation (diamonds) with the number of candidates 10, and the shift method with $c=0.4$ and $w=0.05$ (squares).
  The horizontal axis $\sigma_2 / \sigma_1$ corresponds to the sampling difficulty.
  The statistical errors are in the same order as the point sizes.}
  \label{fig:gaussian-corr-time}
  \end{center}
\end{figure}

Here, as another update method,\cite{suwa_chapter_2012} let us choose the
next state as
\begin{equation}
x'_1 = F^{-1}( \{ F(x_1 | x_2 ) + c + wu \} ),
\end{equation}
where $x_1$ is the current state, $c$ and $w$ are positive real parameters with $c \geq w$, and $u$ is an uniformly random variable in $[-1,1]$, respectively.
The symbol $\{a\}$ takes the fractional portion of a real number $a$.
If we use $w=1/2$, this process is nothing but the Gibbs sampler.
On the other hand, when $w < 1/2$ and ${c} \neq 0, 1/2$, it does not satisfy the detailed balance, and there is a net stochastic flow.
This flow can push the configuration globally, as shown in \cref{fig:gaussian-shift}.
As a result, the autocorrelation time of $(x_1+x_2)^2$ is significantly reduced, as shown in \cref{fig:gaussian-corr-time}.
In this figure, the Gibbs sampler, the over-relaxation methods with $\alpha = - 0.86$, the ordered over-relaxation method (with 10 candidates), and the nonreversible update method with $c=0.4$ and $w=0.05$ are tested.
The nonreversible kernel produces the smallest correlation for $\sigma_2 / \sigma_1 \geq 50$ and achieves about 50 times as short the correlation time as the Gibbs sampler.
We can surely find better parameter sets of the shift algorithm than the best parameter of the conventional over-relaxation methods in almost the whole region.

\begin{figure}[tbp]
  \begin{center}
  \includegraphics[width=0.5\columnwidth]{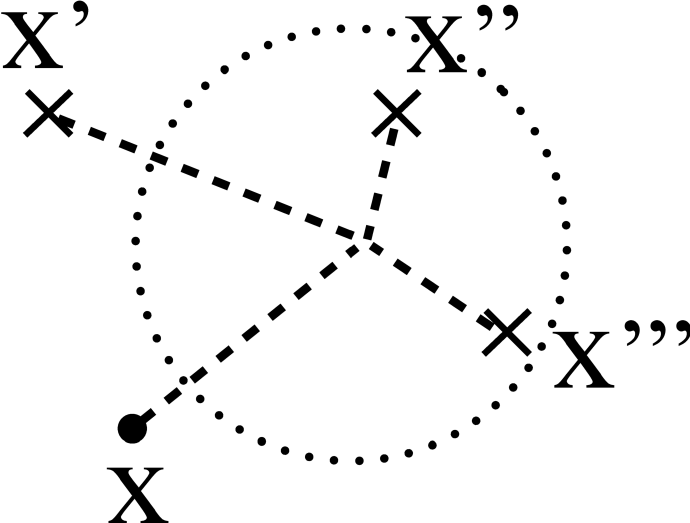}
  \caption{Multiple-proposal example for $n=4$.
  At first, a hub (pivot) is chosen from the current position $X$.
  Then, candidates $X'$, $X''$, and $X'''$ are generated from the hub.
  The dotted line shows the one-sigma line of the Gaussian distribution as a proposal example.}
  \label{fig:multi-proposal}
  \end{center}
\end{figure}

\section{Multiple proposals}
\label{sec:multi-proposal}

In this section, we explain that it is possible to significantly reduce the rejection rate for general cases.
When the direct inversion method, discussed in Sec.~\ref{sec:ags}, cannot be applied, we typically resort to the Metropolis algorithm, where a candidate is generated and accepted/rejected according to the weight and proposal probability ratio.
It has been a canonical MCMC method algorithm since its invention in 1953 (see Ref.~\onlinecite{metropolis_equation_1953}).
However, the inevitable rejection often obstructs efficient sampling.
When the number of candidates is two (including the current state), the Metropolis algorithm achieves a minimized rejection rate.
To reduce the rejection rate, we must prepare more candidates.
Several methods have been proposed as an alternative to the simple Metropolis algorithm.
An example is the multipoint Metropolis method; after generating several candidate states, the next configuration is chosen stochastically with the detailed balance kept.\cite{liu_monte_2004,tjelmeland_using_2004}

After creating a set of candidate states, we can apply the rejection-minimized algorithm.\cite{suwa_markov_2010}
Let us consider sampling from the wine-bottle (Mexican-hat) potential,
\begin{equation}
  \begin{split}
  &\pi(x_1, x_2 ) \\
  &\propto e^{- \left( \frac{ (x_1 - x_2 )^2}{2 \sigma_1^2} + \frac{ ( x_1 + x_2 )^2 }{ 2 \sigma_2^2} \right) \left( \frac{ (x_1 - x_2 )^2}{2 \sigma_1^2} + \frac{ ( x_1 + x_2 )^2 }{ 2 \sigma_2^2} - h \right) + \frac{ h^2}{4}},
  \end{split}
  \label{eqn:wb}
\end{equation}
where $\sigma_1, \sigma_2,$ and $h$ are positive parameters of the potential function.
Then, we propose a candidate configuration using the isotropic bivariate Gaussian distribution $ q(\Delta x_1, \Delta x_2) \propto \exp( -
(\Delta x_1)^2 - (\Delta x_2)^2)$.
As $q$ is symmetric with respect to the current and proposed states, the proposal probability does not need to be taken into account in the global balance condition.

Here, we try to make multiple proposals.
If we propose candidates from the current position and naively make a transition matrix (probability), considering only the weight $\pi$, the global balance is broken.
This is because the joint proposal probability of multiple candidates, including the current state, depends on the current state.
We avoid this problem by introducing a hub (pivot) from which candidates are proposed according to $q(\Delta x_1, \Delta x_2)$, as shown in Fig.~\ref{fig:multi-proposal}.
Since $q$ is symmetric with respect to the hub and each proposed candidate, the joint proposal probability of multiple candidates becomes the same regardless of the current state.
In particular, we use the following multiple-proposal strategy for $n$ candidates.\cite{murray_advances_2007}
\begin{enumerate}
  \item A configuration is chosen as a hub (pivot) from the current configuration by a proposal distribution.
  \item $(n-1)$ candidates are generated from the hub using the same proposal distribution as process 1.
  \item The next state is chosen among the $n$ candidates (including the current state) using the transition probabilities, only taking into account the weights of the states.
\end{enumerate}
This procedure example for $n=4$ is shown in \cref{fig:multi-proposal}.
In process 3, we can apply the shift method and minimize the rejection probability, as explained in Sec.~\ref{sec:ags}.

\begin{figure}[tbp]
  \begin{center}
  \includegraphics[width=1.0\columnwidth]{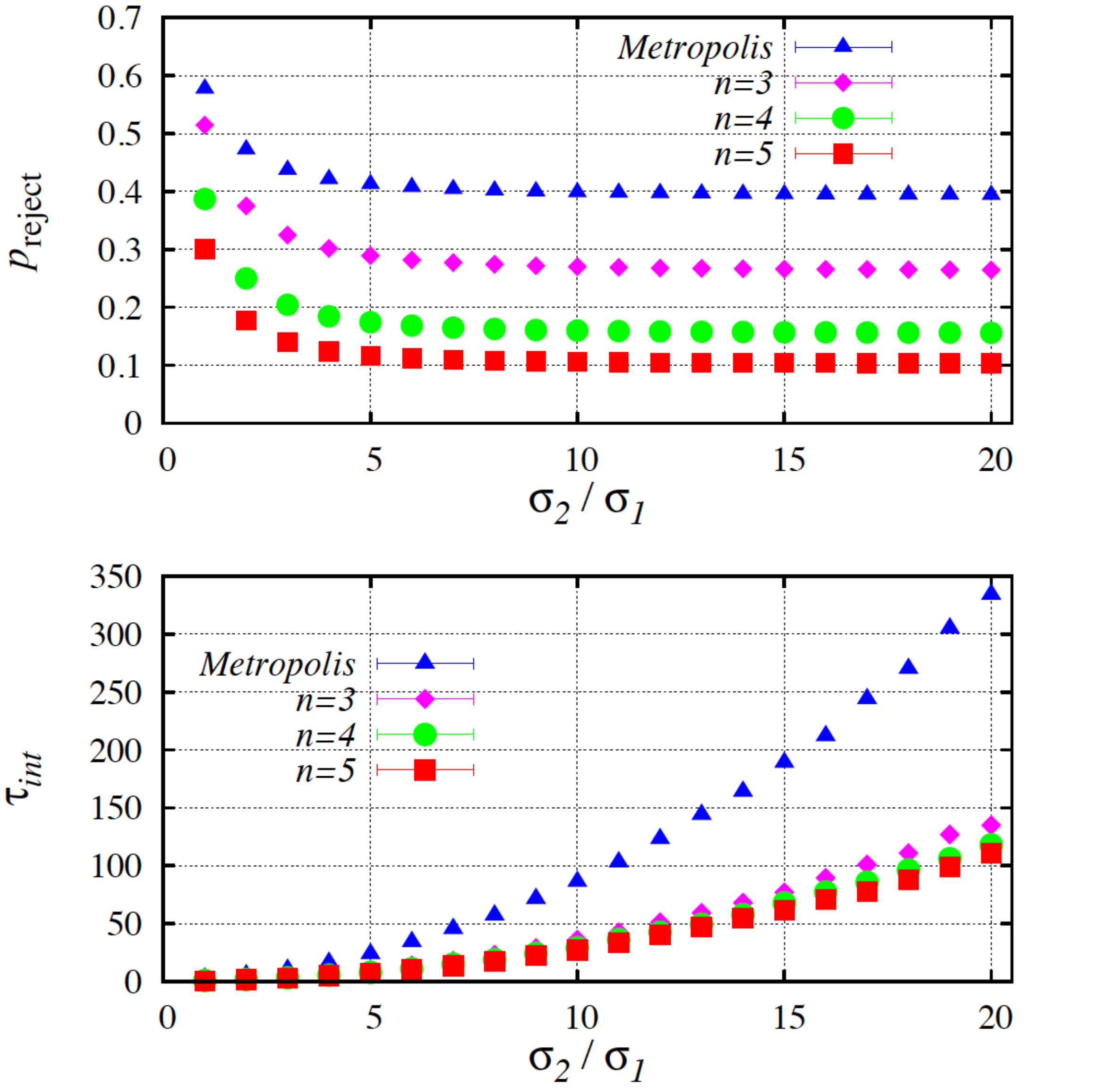}
  \caption{Rejection rates (upper) and the correlation times of $(x_1+x_2)^2$ (lower) from the simple Metropolis algorithm and the rejection-minimized method for $n=3,4,5$ in the wine-bottle potential~[\cref{eqn:wb}] with $h=16$.
  The rejection rate is reduced as the number of candidates is increased.
  Accompanying the rejection rate, the correlation time gets shorter.}
\label{fig:rejection-rate}
\end{center}
\end{figure}

\Cref{fig:rejection-rate} shows that the rejection rate is indeed reduced by using this multiple-proposal algorithm and the nonreversible kernel with the choice of $s=\max_i \pi_i$ as in the Suwa--Todo algorithm.~\cite{suwa_markov_2010}
The correlation time of $(x_1 + x_2)^2$ also gets shorter as the number of candidates is increased.
The extension to higher-dimensional cases is straightforward.

We note that parallel computation can significantly help reduce the computational cost of the multiple-proposal method.
Using a single core, the computational cost in each Monte Carlo update increases proportionally to $n$ for large $n$.
However, candidate states can be prepared in parallel using multicores, and the computation time can be kept almost independent of $n$.
Then, the total computation time needed, the product of the correlation time, and the wall clock time for each Monte Carlo step, decreases with increasing $n$.
Such parallel computation helps simulate hardly relaxing problems, such as protein folding.

\section{Lifting}
\label{sec:lifting}

Proposing good candidate states is crucial for efficient Monte Carlo sampling.
To reduce correlation times, efficient transitions between states with high weights are necessary.
However, since the topology of these high-weight states in state space is usually unknown, except for trivial cases, proposing good candidate states is difficult.

An effective idea is to expand the state space and connect high-weight states with relatively easy paths.
A typical method using this idea is the hybrid (Hamiltonian) Monte Carlo method, which assigns virtual momenta to state variables and updates states according to Newtonian dynamics, enabling efficient transitions between states with constant total energy.\cite{duane_hybrid_1987}
The worm algorithm, introduced later, applies to systems with conservation laws, expanding the state space to one where these laws are broken.

Originally, these algorithms were designed to satisfy the detailed balance.
The lifting technique intentionally expands the state space and introduces probability flow in the expanded space.
Skillfully designed nonreversible MCMC methods can significantly improve sampling efficiency.

\begin{figure}[tbp]
  \begin{center}
  \includegraphics[width=0.97\columnwidth]{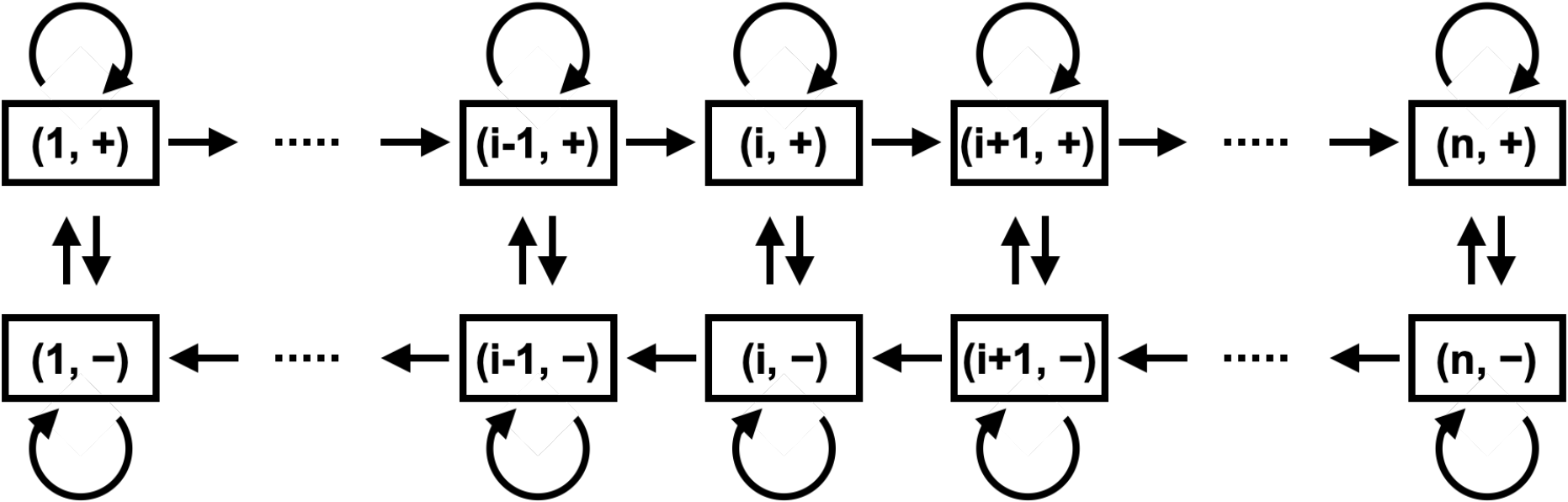}
  \caption{Introducing probability flow through lifting.
  Each of the \( n \) states in one dimension is lifted to two states, \( + \) and \( - \).
  Probability flow is introduced so that \( + \) states move in the positive direction and \( - \) states move in the negative direction.
  When the weights of each state are the same, the mixing time \( \tau_{\text{mix}} \) to traverse the entire system is \( O(n^2) \) without probability flow, whereas introducing flow can reduce this to \( O(n) \).}
  \label{fig:lifting}
  \end{center}
\end{figure}

The clear idea of lifting was proposed by Diaconis et al., who discussed sampling from $n$ states arranged in one dimension, where only one neighboring state can be moved to each time step.\cite{diaconis_analysis_2000}
Without bias, the dynamics are diffusive, and the mixing (relaxation) time is $\tau_{\text{mix}} = O(n^2)$.
Expanding the state space by preparing two states, $+$ and $-$, for each state, and creating probability flow such that $+$ states move in the positive direction and $-$ states in the negative direction, as shown in \cref{fig:lifting}, the global balance is maintained using skewed detailed balance,\cite{turitsyn_irreversible_2011}
\begin{equation}
\pi_i P_{(i, +) \to (i+1, +)} = \pi_{i+1} P_{(i+1, -) \to (i, -)}.
\label{eq:skewed}
\end{equation}
When such a probability flow is introduced, and the transition probabilities are optimally chosen, the time to traverse the entire system can be reduced to $\tau_{\text{mix}} = O(n)$, a ballistic dynamics compared to diffusive dynamics without flow.\cite{diaconis_analysis_2000}
Probability flow can generally shorten relaxation times by up to a square root.
For two-dimensional uniform distributions, similar relaxation time reduction is possible.\cite{chen_lifting_1999,vucelja_liftingnonreversible_2016}
In the mean-field (fully connected) Ising model, state variables are essentially represented by the total energy.
Introducing probability flow reduces the relaxation time at the critical temperature from $\tau_{\text{exp}} \propto N^{3/2}$ to $N^{3/4}$ (see Refs.~\onlinecite{turitsyn_irreversible_2011,fernandes_non-reversible_2011}).
A similar reduction is achieved in one-dimensional Ising and Potts models.\cite{sakai_dynamics_2013,Faizi2020}

One of the successful applications of lifting is the event-chain Monte Carlo (ECMC) method, which performs a rejection-free state update in interacting particle systems in continuous space.~\cite{bernard_event-chain_2009, michel_generalized_2014, krauth_event-chain_2021} 
The ECMC method was first applied to the two-dimensional hard-sphere model, demonstrating that equilibrium could be achieved even in systems with one million particles.~\cite{bernard_event-chain_2009, Engel13, Isobe15}
This simulation clarified phase transition phenomena in two-dimensional systems, which had been debated for decades, including that the transition from the liquid phase to the hexatic phase is a Mayer--Wood type first-order transition, and the transition from the hexatic phase to the solid phase is a Berezinskii--Kosterlitz--Thouless transition. 
The method has also been extended to particle systems with many-body interactions involving three or more bodies~\cite{Harland17} and long-range interactions.~\cite{Kapfer16}
In many cases, it provides more efficient sampling than molecular dynamics in single-threaded computations. 
Furthermore, the ECMC method can be applied not only to particle systems but also to classical spin models. 
For instance, in the three-dimensional classical Heisenberg spin system, the dynamic critical exponent $z$ decreases to one at the critical temperature.~\cite{Nishikawa2015}
In addition, in the low-temperature phase of the three-dimensional XY spin glass, relaxation is accelerated compared to other Monte Carlo methods,~\cite{Michel2015} demonstrating its effectiveness in various systems.~\cite{krauth_event-chain_2021}

\section{Directed Worm Algorithm}

We introduce another application of lifting to statistical mechanics problems, the directed worm algorithm.
Initially proposed for quantum systems with constraints, such as particle number conservation, the worm algorithm extends the state space to include states violating these constraints.\cite{prokofev_exact_1998,boninsegni_worm_2006}
In the worldline representation, it efficiently samples paths that do not break particle conservation.
However, local state updates while maintaining constraints are inefficient, and it is challenging to update topological quantities such as the winding number of worldlines.
Generally, sampling constrained state spaces is a nontrivial problem.

The worm algorithm inserts a pair of kinks that break constraints and performs state updates by using the random walk of these kinks.
By traversing states that violate the constraints, it enables efficient updates and nonlocal state transitions, resolving the problem of sampling topological quantities.
The kinks' random walk resembles a worm, giving the algorithm its name.

The computational efficiency of the worm algorithm depends heavily on the behavior of the kinks' random walk.
As discussed in Sec.~\ref{sec:lifting}, minimizing randomness and achieving ballistic movement improves efficiency.
The idea of lifting is applied to the kinks' random walk by extending the state space to include directional degrees of freedom.
Creating probability flow in this expanded space enhances sampling efficiency.
The improved directed worm algorithm has become a standard method for quantum spins and boson systems.\cite{syljuasen_quantum_2002}

The worm algorithm also applies to many classical systems.
For instance, the Ising model can be reformulated as a constrained problem by converting spin variables into bond variables.\cite{prokofev_worm_2001}
This constraint is naturally suited to the worm algorithm, making it the most efficient computational method for the Ising model.
This reformulation is applicable to models such as the Potts and $\phi^4$ models.\cite{prokofev_worm_2001}
The worm algorithm is widely used in spin glass models,\cite{wang_worm_2005} \(O(n)\) loop models,\cite{liu_worm_2011} and lattice QCD,\cite{adams_chiral_2003} significantly reducing correlation times and becoming one of the most efficient state update methods.
Here, we explain the recently proposed directed worm algorithm for classical systems.\cite{suwa_geometric_2021,suwa_lifted_2022}

\begin{figure}[tbp]
  \begin{center}
  \includegraphics[width=0.97\columnwidth]{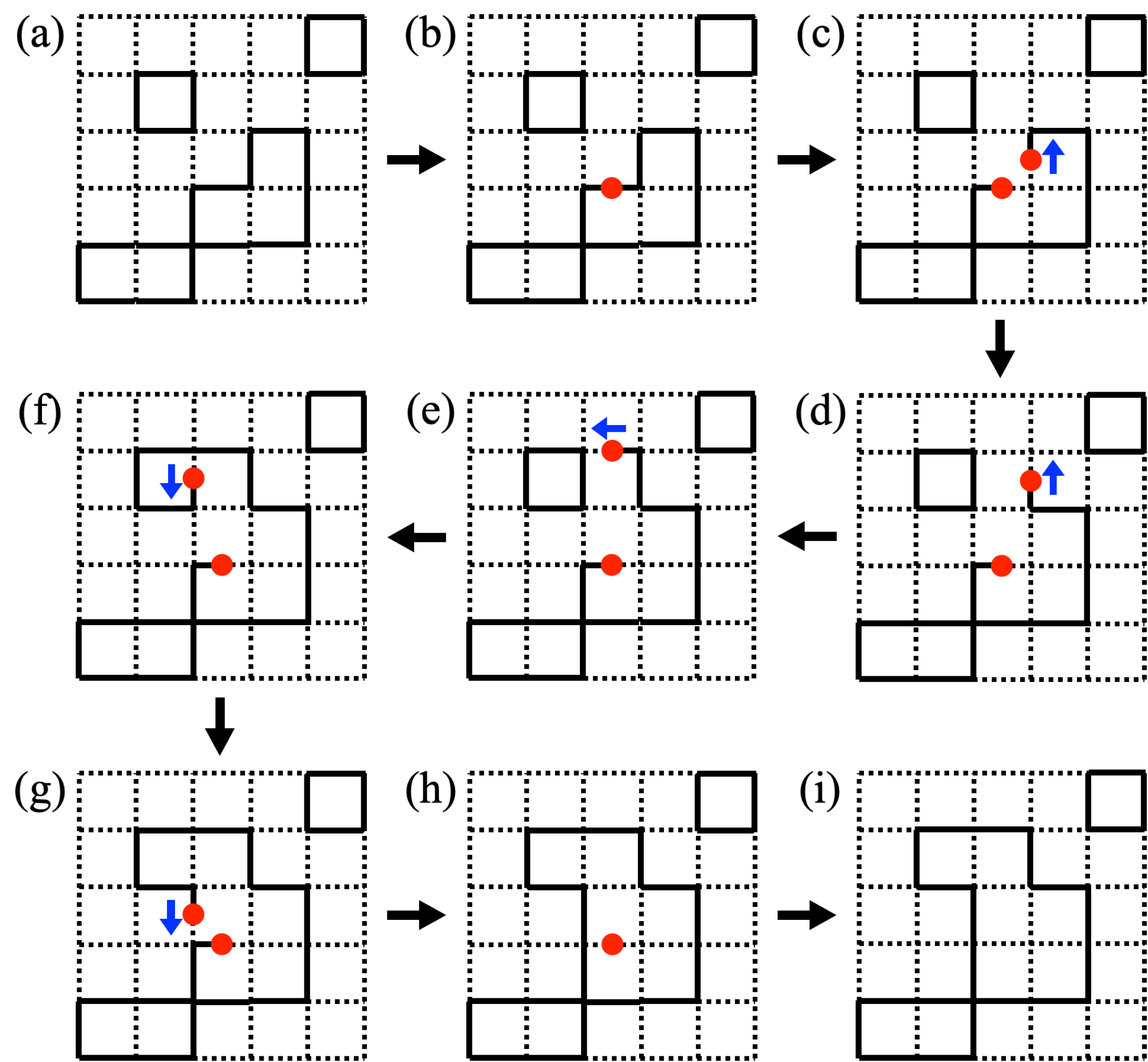}
  \caption{
  Example of updating the directed worm algorithm in two dimensions.
  (a) The active bonds (black solid lines) satisfy the loop constraint.
  (b) A bond is randomly selected, and a pair of kinks (red dots) is inserted.
  (c)--(g) The kinks perform a random walk while having a direction (blue arrows), updating the bond variables.
  (h) The process stops when the two kinks meet again.
  (i) The kinks are removed, and the system transitions to another state satisfying the loop constraint.
  }
  \label{fig:dwa}
\end{center}
\end{figure}

Consider the Ising model on a bipartite lattice with nearest-neighbor interactions.
Let the number of sites be $N$ and the number of bonds be $N_b$.
Transform the site spin variables $\sigma_i = \pm 1$ to bond variables $n_b = 0, 1$.
Defining the spin coupling constant as $J$ and $K = \beta J$, the partition function can be rewritten as
\begin{equation}
  \begin{split}
  Z &= \sum_{\{\sigma_i\}} e^{K \sum_{\langle i,j \rangle} \sigma_i \sigma_j} \\
  &= (2 \cosh K)^{N_b} \sum_{\{n_b\}} \prod_{\langle i,j \rangle} (\tanh K)^{n_b},  
  \end{split}
  \label{eq:Z}
\end{equation}
where the active bonds with $n_b=1$ form loops in the state space [\cref{fig:dwa}(a), (i)].
Nonloop states do not contribute to the partition function.
This representation of the partition function is physically associated with the high-temperature expansion.~\cite{Waerden1941}

Prokof'ev and Svistunov proposed the worm algorithm for systems with loop constraints, inserting kinks, and updating bond variables while the kinks' random-walk.\cite{prokofev_worm_2001}
The transition probabilities are determined by the state weights in the new ensemble.
The directed worm algorithm, introduced by one of the present authors, extends the state space further to include kink directionality.\cite{suwa_geometric_2021}
Kinks are inserted on bonds instead of sites, allowing them to have directional freedom.

The directed worm algorithm follows these steps:
\begin{enumerate}
    \item Randomly select a bond and insert a pair of kinks [\cref{fig:dwa}(b)].
    \item One kink random-walks along bonds with directional freedom, updating bond variables [\cref{fig:dwa}(c)-(g)].
    \item When the kinks meet again, remove them [\cref{fig:dwa}(h) and \cref{fig:dwa}(i)].
\end{enumerate}

We extend the state space by introducing another state for each bond variable so that the variable of the bond on which a kink is located takes $n_b=1/2$.
The important feature of the worm algorithm is that the kink distance corresponds to the system's correlation length, enabling efficient updates of regions of this size.
This feature allows the method to adaptively achieve efficient state updates.

\begin{figure}[tbp]
  \begin{center}
  \includegraphics[width=0.97\columnwidth]{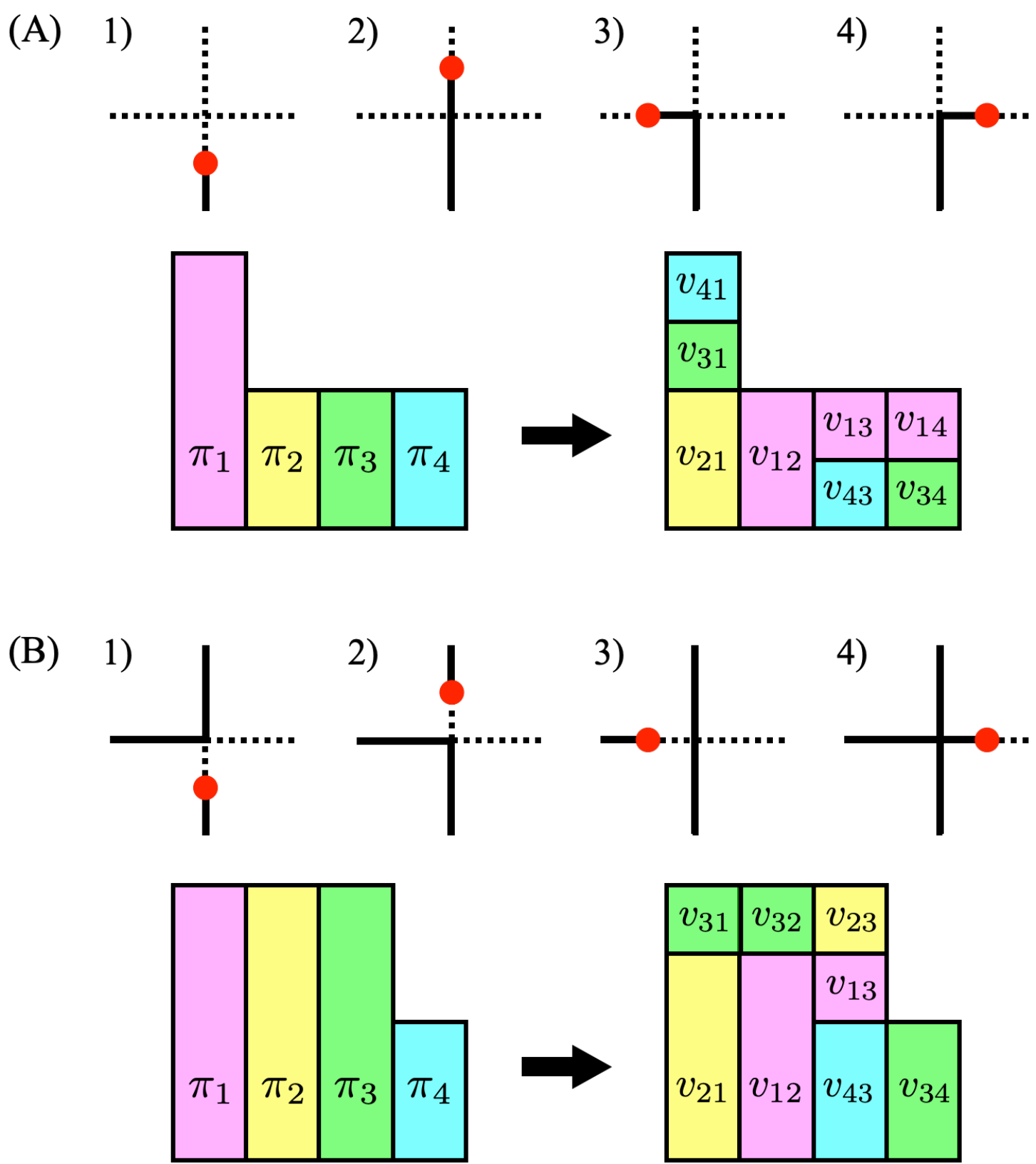}
  \caption{Optimized transition probabilities for the directed worm algorithm in two-dimensional systems using geometric allocation methods.
  The probability of the kink reflecting back to the original bond (rejection rate) is minimized.
  Furthermore, under this condition, the probability of the kink moving straight is maximized.
  On a square lattice, this results in either case (A) or (B) using rotation and reflection.
  Here, the weight ratio is $\pi_4/\pi_1=\tanh K$ according to \cref{eq:Z}.
  These transition probabilities satisfy the skewed detailed balance ($v_{ij}=v_{ji}$).
  (A) For example, the transition probabilities in the case of \cref{fig:dwa}(d).
  With $v_{12}=\pi_4$, $v_{13}=v_{14}=\frac{1}{2}(\pi_1 - \pi_4)$, and $v_{34}=\frac{1}{2}(3\pi_4 - \pi_1)$.
  (B) For example, the transition probabilities in the case of \cref{fig:dwa}(e) (rotated 90$^{\circ}$ clockwise).
  With $v_{12}=\frac{1}{2}(\pi_1 + \pi_4)$, $v_{13}=v_{23}=\frac{1}{2}(\pi_1 - \pi_4)$, and $v_{34}=\pi_4$.
  Partially modified from Ref.~\onlinecite{suwa_geometric_2021}.}
  \label{fig:ga}
  \end{center}
\end{figure}

Optimizing the transition probabilities in the kink's state update is essential to maximize the utility of lifting.
Even after introducing a degree of freedom in orientation, computational efficiency is reduced if there is a process where the kink immediately returns to the path that it has followed.
For optimizing such transition probabilities, the geometric allocation method developed in Ref.~\onlinecite{suwa_markov_2010} is very useful.
This method allows for the intuitive design of probability flows and allows optimizations, such as rejection minimization, to be performed.\cite{todo_geometric_2013}

Consider the case where there are $n$ candidate states to transition to.
If we define a stochastic flow $v_{ij}=\pi_i P_{ij}$, the two conditions that the stochastic flow must satisfy are
\begin{align}
\pi_i &= \sum_{j=1}^n v_{ij} \quad \forall i \label{eq:pc}, \\
\pi_j &= \sum_{i=1}^n v_{ij} \quad \forall j \label{eq:gb},
\end{align}
corresponding to probability conservation and the global balance conditions, respectively.

In the geometric allocation method, the transition probabilities can be determined intuitively, as shown in \cref{fig:ga}.
In this figure, the transition probabilities of the directed worm algorithm in the two-dimensional system ($n=4$) are optimized.
Determining the probability flow to satisfy Eqs.~(\ref{eq:pc}) and (\ref{eq:gb}) is equivalent to rearranging the colors while preserving the area of each color in the diagram and the overall box shape; this is the heart of this approach.
The probability of rejection $\sum_i v_{ii}$ is the sum of the areas of the colors allocated to the original boxes.
In the allocation shown in \cref{fig:ga}, it can be seen that the transition probability has no rejection ($v_{ii}=0$).
In the $d$-dimensional Ising model, the rejection rate can be set to zero in the temperature region $T \leq \frac{2}{\ln \frac{d}{d-1}}=T_\text{Bethe}$, which is the transition temperature in the Bethe approximation and always higher than the true transition temperature for $d < \infty$ (see Ref.~\onlinecite{suwa_lifted_2022}).
Furthermore, here, under a minimal rejection rate, the probability of a straight transition was allocated to be maximal.
In addition, this transition probability satisfies the skewed detailed balance ($v_{ij}=v_{ji}$) [\cref{eq:skewed}].
However, it is easy to break this condition and further increase the irreversibility.\cite{suwa_chapter_2012,suwa_geometrically_2014}
This intuitive and flexible optimization is a significant property of the geometric allocation method.
As shown in the caption of \cref{fig:ga}, the transition probabilities can be written down analytically and efficiently implemented in programming codes.\cite{suwa_markov_2010}
Similar improvements and optimizations can also be made for higher-dimensional Ising models.\cite{suwa_lifted_2022}

\begin{figure}[tbp]
  \begin{center}
  \includegraphics[width=0.97\columnwidth]{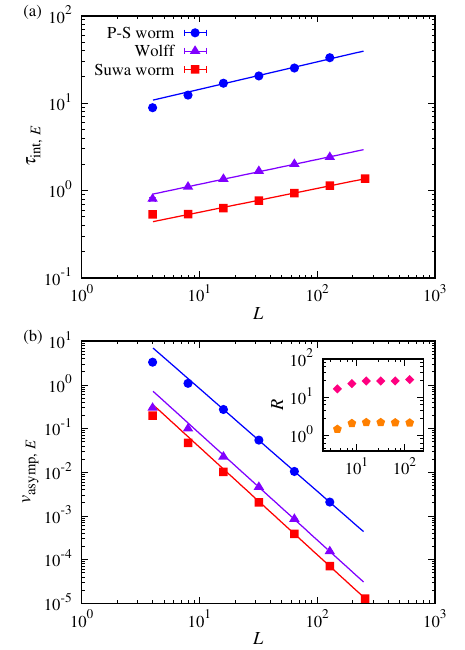}
  \caption{
(a) Integrated autocorrelation time and (b) the asymptotic variance of the total energy as a function of the system length in the simple cubic lattice Ising model at the critical temperature. The Prokof'ev--Svistunov (P--S) worm,~\cite{prokofev_worm_2001} the Wolff cluster,~\cite{wolff_collective_1989} and the directed worm (Suwa)~\cite{suwa_geometric_2021} algorithms are compared. The exponent of $\tau_{{\rm int}, E}$ is estimated to be $z \approx 0.27$ from the result of the directed worm algorithm.
The inset of panel (b) shows the ratios of the asymptotic variance in the P--S worm (diamonds) and the Wolff cluster (pentagons) algorithms to the one in the directed worm algorithm, $\sim$27 and 2.2 for large system sizes, respectively.  
The data were taken from Ref.~\onlinecite{suwa_geometric_2021}.}
  \label{fig:3D-result}
  \end{center}
\end{figure}

In such optimized directed worms, the randomness is reduced, and the kink diffusion coefficient at the critical temperature of the two-dimensional Ising model is six times as large as that of the worm algorithm with a random walk on the site.~\cite{suwa_geometric_2021}
Figure~\ref{fig:3D-result} shows the integrated autocorrelation time and the asymptotic variance of the total energy as a function of the system length at the critical temperature of the three-dimensional Ising model.
The directed worm algorithm produces a significantly shorter autocorrelation time than the Prokof'ev--Svistunov worm~\cite{prokofev_worm_2001} and the Wolff cluster~\cite{wolff_collective_1989} algorithms.
Although the dynamic critical exponent is presumably the same in these algorithms, the sampling efficiency of the directed worm algorithm, measured by the asymptotic variance, is $\sim$27 and 2.2 times as high as that of the Prokof'ev--Svistunov worm and the Wolff cluster algorithms, respectively.
Similar efficiency enhancement is confirmed for the susceptibility.~\cite{suwa_geometric_2021}
It should be noted that the dynamic critical exponent of the Wolff algorithm is estimated to be $z=0.24(2)$ in the three-dimensional Ising model.\cite{Liu2014}
In addition, at the critical temperature of the four-dimensional Ising model, the asymptotic variances of the energy and susceptibility are reduced by a factor of 80.
The directed worm algorithm is more efficient than the widely used Wolff cluster algorithm, one of the most efficient Monte Carlo update methods.
The dynamic critical exponents of the three- and four-dimensional Ising models in the worm algorithm are $z \approx 0.27$ and 0, respectively, significantly smaller than in the usual local state update ($z \approx 2$) (See Ref.~\onlinecite{suwa_geometric_2021} and \onlinecite{suwa_lifted_2022}). 
Therefore, the worm algorithm significantly reduces the dynamic critical exponents of various models, making it a powerful numerical method for studying critical phenomena.

\section{Future Prospects}

Finally, we discuss future prospects for MCMC methods that design probability flow.
Currently, molecular dynamics methods are often used for particle systems in continuous space, but slow relaxation due to interactions remains a major issue.
The ECMC method is expected to contribute significantly to sampling from the Boltzmann distribution of systems.\cite{bernard_event-chain_2009,michel_generalized_2014,krauth_event-chain_2021}
However, it is challenging to restrict the probability processes to local regions, making parallel computation difficult.
Recently, shared-memory parallel computation methods for ECMC have been developed, accelerating calculations by about ten times compared to single-threaded computations.\cite{li_multithreaded_2021}
Developing distributed-memory parallel computation methods could lead to a significant breakthrough.

The directed worm algorithm is applicable not only to the Ising model but also to many quantum and classical models.
As another application of lifting, a worm algorithm introducing probability flow along the energy axis has been proposed.\cite{elci_lifted_2018}
The fusion of these methods and the introduction of different types of probability flow are expected to lead to further improvements.\cite{suwa_lifted_2022}

In addition to the algorithms introduced in this paper, MCMC methods introducing probability flow are used in various contexts.
For example, simulated tempering and replica exchange methods can significantly reduce relaxation times by using lifting.\cite{sakai_irreversible_2016,syed_non-reversible_2022}
Optimizing replica exchange probabilities by breaking detailed balance with geometric allocation methods can further enhance efficiency.\cite{itoh_replica-permutation_2013}
These attempts to introduce probability flow into extended ensemble methods have broad applications and are important research topics.
In addition, nonreversible Metropolis--Hastings methods introducing probability vortices have been proposed.\cite{bierkens_non-reversible_2016}
The development of various nonreversible MCMC methods is expected to continue.

Not only MCMC methods but also stochastic processes such as Langevin dynamics are gaining attention for dynamics that do not satisfy detailed balance.
In nonreversible Langevin dynamics, net probability flow accelerates distribution convergence and reduces asymptotic variance.\cite{ohzeki_langevin_2015,duncan_variance_2016}

Nonequilibrium steady states are being actively researched not only for sampling but also for extending equilibrium statistical mechanics and studying open systems.\cite{ichiki_violation_2013,coghi_role_2021}
Mathematically, convergence and ergodicity of nonreversible Markov chains are major research themes.\cite{abdulle_accelerated_2019,wang_regeneration-enriched_2021}
These new developments are expected to form the foundational theory for nonreversible MCMC methods and guide the development of efficient algorithms.

\begin{acknowledgments}
This work was supported by JSPS KAKENHI under Grant Nos. JP20H01824, JP22K03508, JP24H01609, and JP24K00543, by the Center of Innovation for Sustainable Quantum AI (SQAI), JST under Grant No. JPMJPF2221, and by CREST under Grant No. JPMJCR24I1.
\end{acknowledgments}

\section*{References}
\bibliography{main}

\begin{thebibliography}{60}%
\makeatletter
\providecommand \@ifxundefined [1]{%
 \@ifx{#1\undefined}
}%
\providecommand \@ifnum [1]{%
 \ifnum #1\expandafter \@firstoftwo
 \else \expandafter \@secondoftwo
 \fi
}%
\providecommand \@ifx [1]{%
 \ifx #1\expandafter \@firstoftwo
 \else \expandafter \@secondoftwo
 \fi
}%
\providecommand \natexlab [1]{#1}%
\providecommand \enquote  [1]{``#1''}%
\providecommand \bibnamefont  [1]{#1}%
\providecommand \bibfnamefont [1]{#1}%
\providecommand \citenamefont [1]{#1}%
\providecommand \href@noop [0]{\@secondoftwo}%
\providecommand \href [0]{\begingroup \@sanitize@url \@href}%
\providecommand \@href[1]{\@@startlink{#1}\@@href}%
\providecommand \@@href[1]{\endgroup#1\@@endlink}%
\providecommand \@sanitize@url [0]{\catcode `\\12\catcode `\$12\catcode
  `\&12\catcode `\#12\catcode `\^12\catcode `\_12\catcode `\%12\relax}%
\providecommand \@@startlink[1]{}%
\providecommand \@@endlink[0]{}%
\providecommand \url  [0]{\begingroup\@sanitize@url \@url }%
\providecommand \@url [1]{\endgroup\@href {#1}{\urlprefix }}%
\providecommand \urlprefix  [0]{URL }%
\providecommand \Eprint [0]{\href }%
\providecommand \doibase [0]{http://dx.doi.org/}%
\providecommand \selectlanguage [0]{\@gobble}%
\providecommand \bibinfo  [0]{\@secondoftwo}%
\providecommand \bibfield  [0]{\@secondoftwo}%
\providecommand \translation [1]{[#1]}%
\providecommand \BibitemOpen [0]{}%
\providecommand \bibitemStop [0]{}%
\providecommand \bibitemNoStop [0]{.\EOS\space}%
\providecommand \EOS [0]{\spacefactor3000\relax}%
\providecommand \BibitemShut  [1]{\csname bibitem#1\endcsname}%
\let\auto@bib@innerbib\@empty
\bibitem [{\citenamefont {Newman}\ and\ \citenamefont
  {Barkema}(1999)}]{newman_monte_1999}%
  \BibitemOpen
  \bibfield  {author} {\bibinfo {author} {\bibfnamefont {M.}~\bibnamefont
  {Newman}}\ and\ \bibinfo {author} {\bibfnamefont {G.}~\bibnamefont
  {Barkema}},\ }\href@noop {} {\emph {\bibinfo {title} {Monte {Carlo} methods
  in statistical physics}}}\ (\bibinfo  {publisher} {Oxford University Press},\
  \bibinfo {year} {1999})\BibitemShut {NoStop}%
\bibitem [{\citenamefont {Landau}\ and\ \citenamefont
  {Binder}(2014)}]{landau_guide_2014}%
  \BibitemOpen
  \bibfield  {author} {\bibinfo {author} {\bibfnamefont {D.~P.}\ \bibnamefont
  {Landau}}\ and\ \bibinfo {author} {\bibfnamefont {K.}~\bibnamefont
  {Binder}},\ }\href
  {https://www.cambridge.org/core/books/guide-to-monte-carlo-simulations-in-statistical-physics/2522172663AF92943C625056C14F6055}
  {\emph {\bibinfo {title} {A {Guide} to {Monte} {Carlo} {Simulations} in
  {Statistical} {Physics}}}},\ \bibinfo {edition} {4th}\ ed.\ (\bibinfo
  {publisher} {Cambridge University Press},\ \bibinfo {address} {Cambridge},\
  \bibinfo {year} {2014})\BibitemShut {NoStop}%
\bibitem [{\citenamefont {Sakai}\ and\ \citenamefont
  {Hukushima}(2013)}]{sakai_dynamics_2013}%
  \BibitemOpen
  \bibfield  {author} {\bibinfo {author} {\bibfnamefont {Y.}~\bibnamefont
  {Sakai}}\ and\ \bibinfo {author} {\bibfnamefont {K.}~\bibnamefont
  {Hukushima}},\ }\href {\doibase 10.7566/JPSJ.82.064003} {\bibfield  {journal}
  {\bibinfo  {journal} {J. Phys. Soc. Jpn.}\ }\textbf {\bibinfo {volume}
  {82}},\ \bibinfo {pages} {064003} (\bibinfo {year} {2013})}\BibitemShut
  {NoStop}%
\bibitem [{\citenamefont {Faizi}, \citenamefont {Deligiannidis},\ and\
  \citenamefont {Rosta}(2020)}]{Faizi2020}%
  \BibitemOpen
  \bibfield  {author} {\bibinfo {author} {\bibfnamefont {F.}~\bibnamefont
  {Faizi}}, \bibinfo {author} {\bibfnamefont {G.}~\bibnamefont
  {Deligiannidis}}, \ and\ \bibinfo {author} {\bibfnamefont {E.}~\bibnamefont
  {Rosta}},\ }\href {\doibase 10.1021/acs.jctc.9b01135} {\bibfield  {journal}
  {\bibinfo  {journal} {Journal of Chemical Theory and Computation}\ }\textbf
  {\bibinfo {volume} {16}},\ \bibinfo {pages} {2124} (\bibinfo {year}
  {2020})}\BibitemShut {NoStop}%
\bibitem [{\citenamefont {Nishikawa}\ \emph {et~al.}(2015)\citenamefont
  {Nishikawa}, \citenamefont {Michel}, \citenamefont {Krauth},\ and\
  \citenamefont {Hukushima}}]{Nishikawa2015}%
  \BibitemOpen
  \bibfield  {author} {\bibinfo {author} {\bibfnamefont {Y.}~\bibnamefont
  {Nishikawa}}, \bibinfo {author} {\bibfnamefont {M.}~\bibnamefont {Michel}},
  \bibinfo {author} {\bibfnamefont {W.}~\bibnamefont {Krauth}}, \ and\ \bibinfo
  {author} {\bibfnamefont {K.}~\bibnamefont {Hukushima}},\ }\href {\doibase
  10.1103/PhysRevE.92.063306} {\bibfield  {journal} {\bibinfo  {journal} {Phys.
  Rev. E}\ }\textbf {\bibinfo {volume} {92}},\ \bibinfo {pages} {063306}
  (\bibinfo {year} {2015})}\BibitemShut {NoStop}%
\bibitem [{\citenamefont {Metropolis}\ \emph {et~al.}(1953)\citenamefont
  {Metropolis}, \citenamefont {Rosenbluth}, \citenamefont {Rosenbluth},
  \citenamefont {Teller},\ and\ \citenamefont
  {Teller}}]{metropolis_equation_1953}%
  \BibitemOpen
  \bibfield  {author} {\bibinfo {author} {\bibfnamefont {N.}~\bibnamefont
  {Metropolis}}, \bibinfo {author} {\bibfnamefont {A.~W.}\ \bibnamefont
  {Rosenbluth}}, \bibinfo {author} {\bibfnamefont {M.~N.}\ \bibnamefont
  {Rosenbluth}}, \bibinfo {author} {\bibfnamefont {A.~H.}\ \bibnamefont
  {Teller}}, \ and\ \bibinfo {author} {\bibfnamefont {E.}~\bibnamefont
  {Teller}},\ }\href {\doibase 10.1063/1.1699114} {\bibfield  {journal}
  {\bibinfo  {journal} {J. Chem. Phys.}\ }\textbf {\bibinfo {volume} {21}},\
  \bibinfo {pages} {1087} (\bibinfo {year} {1953})}\BibitemShut {NoStop}%
\bibitem [{\citenamefont {Creutz}(1980)}]{creutz_monte_1980}%
  \BibitemOpen
  \bibfield  {author} {\bibinfo {author} {\bibfnamefont {M.}~\bibnamefont
  {Creutz}},\ }\href {\doibase 10.1103/PhysRevD.21.2308} {\bibfield  {journal}
  {\bibinfo  {journal} {Phys. Rev. D}\ }\textbf {\bibinfo {volume} {21}},\
  \bibinfo {pages} {2308} (\bibinfo {year} {1980})}\BibitemShut {NoStop}%
\bibitem [{\citenamefont {Geman}\ and\ \citenamefont
  {Geman}(1984)}]{geman_stochastic_1984}%
  \BibitemOpen
  \bibfield  {author} {\bibinfo {author} {\bibfnamefont {S.}~\bibnamefont
  {Geman}}\ and\ \bibinfo {author} {\bibfnamefont {D.}~\bibnamefont {Geman}},\
  }\href {\doibase 10.1109/TPAMI.1984.4767596} {\bibfield  {journal} {\bibinfo
  {journal} {IEEE Trans. Pattern Anal. Mach. Intell.}\ ,\ \bibinfo {pages}
  {721}} (\bibinfo {year} {1984})}\BibitemShut {NoStop}%
\bibitem [{\citenamefont {Robert}\ and\ \citenamefont
  {Casella}(2004)}]{robert_monte_2004}%
  \BibitemOpen
  \bibfield  {author} {\bibinfo {author} {\bibfnamefont {C.~P.}\ \bibnamefont
  {Robert}}\ and\ \bibinfo {author} {\bibfnamefont {G.}~\bibnamefont
  {Casella}},\ }\href
  {https://link.springer.com/book/10.1007/978-1-4757-4145-2} {\emph {\bibinfo
  {title} {Monte {Carlo} {Statistical} {Methods}}}},\ \bibinfo {edition} {2nd}\
  ed.\ (\bibinfo  {publisher} {Springer},\ \bibinfo {address} {New York},\
  \bibinfo {year} {2004})\BibitemShut {NoStop}%
\bibitem [{Note1()}]{Note1}%
  \BibitemOpen
  \bibinfo {note} {In the present paper, we assume irreducibility of the Markov
  chain, which ensures, under global balance, convergence of the running
  average to the expectation value with respect to the target.}\BibitemShut
  {Stop}%
\bibitem [{\citenamefont {Hwang}, \citenamefont {Hwang-Ma},\ and\ \citenamefont
  {Sheu}(2005)}]{hwang_accelerating_2005}%
  \BibitemOpen
  \bibfield  {author} {\bibinfo {author} {\bibfnamefont {C.-R.}\ \bibnamefont
  {Hwang}}, \bibinfo {author} {\bibfnamefont {S.-Y.}\ \bibnamefont {Hwang-Ma}},
  \ and\ \bibinfo {author} {\bibfnamefont {S.-J.}\ \bibnamefont {Sheu}},\
  }\href {\doibase 10.1214/105051605000000025} {\bibfield  {journal} {\bibinfo
  {journal} {Ann. Appl. Probab.}\ }\textbf {\bibinfo {volume} {15}},\ \bibinfo
  {pages} {1433} (\bibinfo {year} {2005})}\BibitemShut {NoStop}%
\bibitem [{\citenamefont {Ichiki}\ and\ \citenamefont
  {Ohzeki}(2013)}]{ichiki_violation_2013}%
  \BibitemOpen
  \bibfield  {author} {\bibinfo {author} {\bibfnamefont {A.}~\bibnamefont
  {Ichiki}}\ and\ \bibinfo {author} {\bibfnamefont {M.}~\bibnamefont
  {Ohzeki}},\ }\href {\doibase 10.1103/PhysRevE.88.020101} {\bibfield
  {journal} {\bibinfo  {journal} {Phys. Rev. E}\ }\textbf {\bibinfo {volume}
  {88}},\ \bibinfo {pages} {020101} (\bibinfo {year} {2013})}\BibitemShut
  {NoStop}%
\bibitem [{\citenamefont {Duncan}, \citenamefont {Lelièvre},\ and\
  \citenamefont {Pavliotis}(2016)}]{duncan_variance_2016}%
  \BibitemOpen
  \bibfield  {author} {\bibinfo {author} {\bibfnamefont {A.~B.}\ \bibnamefont
  {Duncan}}, \bibinfo {author} {\bibfnamefont {T.}~\bibnamefont {Lelièvre}}, \
  and\ \bibinfo {author} {\bibfnamefont {G.~A.}\ \bibnamefont {Pavliotis}},\
  }\href {\doibase 10.1007/s10955-016-1491-2} {\bibfield  {journal} {\bibinfo
  {journal} {J. Stat. Phys.}\ }\textbf {\bibinfo {volume} {163}},\ \bibinfo
  {pages} {457} (\bibinfo {year} {2016})}\BibitemShut {NoStop}%
\bibitem [{\citenamefont {Sokal}(1997)}]{Sokal1997}%
  \BibitemOpen
  \bibfield  {author} {\bibinfo {author} {\bibfnamefont {A.}~\bibnamefont
  {Sokal}},\ }\href@noop {} {\emph {\bibinfo {title} {{Monte} {Carlo} Methods
  in Statistical Mechanics: Foundations and New Algorithms}}}\ (\bibinfo
  {publisher} {Springer},\ \bibinfo {address} {New York},\ \bibinfo {year}
  {1997})\BibitemShut {NoStop}%
\bibitem [{\citenamefont {Suwa}\ and\ \citenamefont
  {Todo}(2010)}]{suwa_markov_2010}%
  \BibitemOpen
  \bibfield  {author} {\bibinfo {author} {\bibfnamefont {H.}~\bibnamefont
  {Suwa}}\ and\ \bibinfo {author} {\bibfnamefont {S.}~\bibnamefont {Todo}},\
  }\href {\doibase 10.1103/PhysRevLett.105.120603} {\bibfield  {journal}
  {\bibinfo  {journal} {Phys. Rev. Lett.}\ }\textbf {\bibinfo {volume} {105}},\
  \bibinfo {pages} {120603} (\bibinfo {year} {2010})}\BibitemShut {NoStop}%
\bibitem [{\citenamefont {Todo}\ and\ \citenamefont
  {Suwa}(2013)}]{todo_geometric_2013}%
  \BibitemOpen
  \bibfield  {author} {\bibinfo {author} {\bibfnamefont {S.}~\bibnamefont
  {Todo}}\ and\ \bibinfo {author} {\bibfnamefont {H.}~\bibnamefont {Suwa}},\
  }\href {\doibase 10.1088/1742-6596/473/1/012013} {\bibfield  {journal}
  {\bibinfo  {journal} {J. Phys.: Conf. Ser.}\ }\textbf {\bibinfo {volume}
  {473}},\ \bibinfo {pages} {012013} (\bibinfo {year} {2013})}\BibitemShut
  {NoStop}%
\bibitem [{\citenamefont {Suwa}(2024)}]{suwa_reducing_2024}%
  \BibitemOpen
  \bibfield  {author} {\bibinfo {author} {\bibfnamefont {H.}~\bibnamefont
  {Suwa}},\ }\href {\doibase 10.1016/j.physa.2023.129368} {\bibfield  {journal}
  {\bibinfo  {journal} {Phy. A: Stat. Mech. Appl.}\ }\textbf {\bibinfo {volume}
  {633}},\ \bibinfo {pages} {129368} (\bibinfo {year} {2024})}\BibitemShut
  {NoStop}%
\bibitem [{\citenamefont {Adler}(1981)}]{adler_over-relaxation_1981}%
  \BibitemOpen
  \bibfield  {author} {\bibinfo {author} {\bibfnamefont {S.~L.}\ \bibnamefont
  {Adler}},\ }\href {\doibase 10.1103/PhysRevD.23.2901} {\bibfield  {journal}
  {\bibinfo  {journal} {Phys. Rev. D}\ }\textbf {\bibinfo {volume} {23}},\
  \bibinfo {pages} {2901} (\bibinfo {year} {1981})}\BibitemShut {NoStop}%
\bibitem [{\citenamefont {Neal}(1998)}]{neal_suppressing_1998}%
  \BibitemOpen
  \bibfield  {author} {\bibinfo {author} {\bibfnamefont {R.~M.}\ \bibnamefont
  {Neal}},\ }in\ \href {https://doi.org/10.1007/978-94-011-5014-9_8} {\emph
  {\bibinfo {booktitle} {Learning in {Graphical} {Models}}}},\ \bibinfo
  {editor} {edited by\ \bibinfo {editor} {\bibfnamefont {M.~I.}\ \bibnamefont
  {Jordan}}}\ (\bibinfo  {publisher} {Springer Netherlands},\ \bibinfo
  {address} {Dordrecht},\ \bibinfo {year} {1998})\ pp.\ \bibinfo {pages}
  {205--228}\BibitemShut {NoStop}%
\bibitem [{\citenamefont {Suwa}(2014)}]{suwa_geometrically_2014}%
  \BibitemOpen
  \bibfield  {author} {\bibinfo {author} {\bibfnamefont {H.}~\bibnamefont
  {Suwa}},\ }\href {https://link.springer.com/10.1007/978-4-431-54517-0} {\emph
  {\bibinfo {title} {Geometrically {Constructed} {Markov} {Chain} {Monte}
  {Carlo} {Study} of {Quantum} {Spin}-phonon {Complex} {Systems}}}},\ Springer
  {Theses}\ (\bibinfo  {publisher} {Springer Japan},\ \bibinfo {address}
  {Tokyo},\ \bibinfo {year} {2014})\BibitemShut {NoStop}%
\bibitem [{\citenamefont {Suwa}\ and\ \citenamefont
  {Todo}(2012)}]{suwa_chapter_2012}%
  \BibitemOpen
  \bibfield  {author} {\bibinfo {author} {\bibfnamefont {H.}~\bibnamefont
  {Suwa}}\ and\ \bibinfo {author} {\bibfnamefont {S.}~\bibnamefont {Todo}},\
  }in\ \href
  {https://www.degruyter.com/document/doi/10.1515/9783110293586.213/pdf?licenseType=restricted}
  {\emph {\bibinfo {booktitle} {Chapter 23. {Geometric} {Allocation} {Approach}
  for the {Transition} {Kernel} of a {Markov} {Chain}}}}\ (\bibinfo
  {publisher} {De Gruyter},\ \bibinfo {year} {2012})\ pp.\ \bibinfo {pages}
  {213--222}\BibitemShut {NoStop}%
\bibitem [{\citenamefont {Liu}(2004)}]{liu_monte_2004}%
  \BibitemOpen
  \bibfield  {author} {\bibinfo {author} {\bibfnamefont {J.~S.}\ \bibnamefont
  {Liu}},\ }\href {http://link.springer.com/10.1007/978-0-387-76371-2} {\emph
  {\bibinfo {title} {Monte {Carlo} {Strategies} in {Scientific}
  {Computing}}}},\ Springer {Series} in {Statistics}\ (\bibinfo  {publisher}
  {Springer},\ \bibinfo {address} {New York, NY},\ \bibinfo {year}
  {2004})\BibitemShut {NoStop}%
\bibitem [{\citenamefont {Tjelmeland}(2004)}]{tjelmeland_using_2004}%
  \BibitemOpen
  \bibfield  {author} {\bibinfo {author} {\bibfnamefont {H.}~\bibnamefont
  {Tjelmeland}},\ }\href@noop {} {\enquote {\bibinfo {title} {Using all
  {Metropolis}-{Hastings} proposals to estimate mean values},}\ }\bibinfo
  {type} {Tech. Rep.}\ \bibinfo {number} {4}\ (\bibinfo  {institution}
  {Norwegian University of Science and Technology Trondheim},\ \bibinfo
  {address} {Norway},\ \bibinfo {year} {2004})\BibitemShut {NoStop}%
\bibitem [{\citenamefont {Murray}(2007)}]{murray_advances_2007}%
  \BibitemOpen
  \bibfield  {author} {\bibinfo {author} {\bibfnamefont {I.}~\bibnamefont
  {Murray}},\ }\emph {\bibinfo {title} {Advances in {Markov} chain {Monte}
  {Carlo} methods}},\ \href@noop {} {\bibinfo {type} {Ph.{D} thesis}},\
  \bibinfo  {school} {University College London} (\bibinfo {year}
  {2007})\BibitemShut {NoStop}%
\bibitem [{\citenamefont {Duane}\ \emph {et~al.}(1987)\citenamefont {Duane},
  \citenamefont {Kennedy}, \citenamefont {Pendleton},\ and\ \citenamefont
  {Roweth}}]{duane_hybrid_1987}%
  \BibitemOpen
  \bibfield  {author} {\bibinfo {author} {\bibfnamefont {S.}~\bibnamefont
  {Duane}}, \bibinfo {author} {\bibfnamefont {A.~D.}\ \bibnamefont {Kennedy}},
  \bibinfo {author} {\bibfnamefont {B.~J.}\ \bibnamefont {Pendleton}}, \ and\
  \bibinfo {author} {\bibfnamefont {D.}~\bibnamefont {Roweth}},\ }\href
  {\doibase 10.1016/0370-2693(87)91197-X} {\bibfield  {journal} {\bibinfo
  {journal} {Phys. Lett. B}\ }\textbf {\bibinfo {volume} {195}},\ \bibinfo
  {pages} {216} (\bibinfo {year} {1987})}\BibitemShut {NoStop}%
\bibitem [{\citenamefont {Diaconis}, \citenamefont {Holmes},\ and\
  \citenamefont {Neal}(2000)}]{diaconis_analysis_2000}%
  \BibitemOpen
  \bibfield  {author} {\bibinfo {author} {\bibfnamefont {P.}~\bibnamefont
  {Diaconis}}, \bibinfo {author} {\bibfnamefont {S.}~\bibnamefont {Holmes}}, \
  and\ \bibinfo {author} {\bibfnamefont {R.~M.}\ \bibnamefont {Neal}},\ }\href
  {\doibase 10.1214/aoap/1019487508} {\bibfield  {journal} {\bibinfo  {journal}
  {Ann. Appl. Probab.}\ }\textbf {\bibinfo {volume} {10}},\ \bibinfo {pages}
  {726} (\bibinfo {year} {2000})}\BibitemShut {NoStop}%
\bibitem [{\citenamefont {Turitsyn}, \citenamefont {Chertkov},\ and\
  \citenamefont {Vucelja}(2011)}]{turitsyn_irreversible_2011}%
  \BibitemOpen
  \bibfield  {author} {\bibinfo {author} {\bibfnamefont {K.~S.}\ \bibnamefont
  {Turitsyn}}, \bibinfo {author} {\bibfnamefont {M.}~\bibnamefont {Chertkov}},
  \ and\ \bibinfo {author} {\bibfnamefont {M.}~\bibnamefont {Vucelja}},\ }\href
  {\doibase 10.1016/j.physd.2010.10.003} {\bibfield  {journal} {\bibinfo
  {journal} {Phys. D: Nonlinear Phenom.}\ }\textbf {\bibinfo {volume} {240}},\
  \bibinfo {pages} {410} (\bibinfo {year} {2011})}\BibitemShut {NoStop}%
\bibitem [{\citenamefont {Chen}, \citenamefont {Lovász},\ and\ \citenamefont
  {Pak}(1999)}]{chen_lifting_1999}%
  \BibitemOpen
  \bibfield  {author} {\bibinfo {author} {\bibfnamefont {F.}~\bibnamefont
  {Chen}}, \bibinfo {author} {\bibfnamefont {L.}~\bibnamefont {Lovász}}, \
  and\ \bibinfo {author} {\bibfnamefont {I.}~\bibnamefont {Pak}},\ }in\ \href
  {\doibase 10.1145/301250.301315} {\emph {\bibinfo {booktitle} {Proceedings of
  the thirty-first annual {ACM} symposium on {Theory} of {Computing}}}},\
  \bibinfo {series and number} {{STOC} '99}\ (\bibinfo  {publisher}
  {Association for Computing Machinery},\ \bibinfo {address} {New York, NY,
  USA},\ \bibinfo {year} {1999})\ pp.\ \bibinfo {pages} {275--281}\BibitemShut
  {NoStop}%
\bibitem [{\citenamefont {Vucelja}(2016)}]{vucelja_liftingnonreversible_2016}%
  \BibitemOpen
  \bibfield  {author} {\bibinfo {author} {\bibfnamefont {M.}~\bibnamefont
  {Vucelja}},\ }\href {\doibase 10.1119/1.4961596} {\bibfield  {journal}
  {\bibinfo  {journal} {Am. J. Phys.}\ }\textbf {\bibinfo {volume} {84}},\
  \bibinfo {pages} {958} (\bibinfo {year} {2016})}\BibitemShut {NoStop}%
\bibitem [{\citenamefont {Fernandes}\ and\ \citenamefont
  {Weigel}(2011)}]{fernandes_non-reversible_2011}%
  \BibitemOpen
  \bibfield  {author} {\bibinfo {author} {\bibfnamefont {H.~C.~M.}\
  \bibnamefont {Fernandes}}\ and\ \bibinfo {author} {\bibfnamefont
  {M.}~\bibnamefont {Weigel}},\ }\href {\doibase 10.1016/j.cpc.2010.11.017}
  {\bibfield  {journal} {\bibinfo  {journal} {Comput. Phys. Commun.}\ }\bibinfo
  {series} {Computer {Physics} {Communications} {Special} {Edition} for
  {Conference} on {Computational} {Physics} {Trondheim}, {Norway}, {June}
  23-26, 2010},\ \textbf {\bibinfo {volume} {182}},\ \bibinfo {pages} {1856}
  (\bibinfo {year} {2011})}\BibitemShut {NoStop}%
\bibitem [{\citenamefont {Bernard}, \citenamefont {Krauth},\ and\ \citenamefont
  {Wilson}(2009)}]{bernard_event-chain_2009}%
  \BibitemOpen
  \bibfield  {author} {\bibinfo {author} {\bibfnamefont {E.~P.}\ \bibnamefont
  {Bernard}}, \bibinfo {author} {\bibfnamefont {W.}~\bibnamefont {Krauth}}, \
  and\ \bibinfo {author} {\bibfnamefont {D.~B.}\ \bibnamefont {Wilson}},\
  }\href {\doibase 10.1103/PhysRevE.80.056704} {\bibfield  {journal} {\bibinfo
  {journal} {Phys. Rev. E}\ }\textbf {\bibinfo {volume} {80}},\ \bibinfo
  {pages} {056704} (\bibinfo {year} {2009})}\BibitemShut {NoStop}%
\bibitem [{\citenamefont {Michel}, \citenamefont {Kapfer},\ and\ \citenamefont
  {Krauth}(2014)}]{michel_generalized_2014}%
  \BibitemOpen
  \bibfield  {author} {\bibinfo {author} {\bibfnamefont {M.}~\bibnamefont
  {Michel}}, \bibinfo {author} {\bibfnamefont {S.~C.}\ \bibnamefont {Kapfer}},
  \ and\ \bibinfo {author} {\bibfnamefont {W.}~\bibnamefont {Krauth}},\ }\href
  {\doibase 10.1063/1.4863991} {\bibfield  {journal} {\bibinfo  {journal} {J.
  Chem. Phys.}\ }\textbf {\bibinfo {volume} {140}},\ \bibinfo {pages} {054116}
  (\bibinfo {year} {2014})}\BibitemShut {NoStop}%
\bibitem [{\citenamefont {Krauth}(2021)}]{krauth_event-chain_2021}%
  \BibitemOpen
  \bibfield  {author} {\bibinfo {author} {\bibfnamefont {W.}~\bibnamefont
  {Krauth}},\ }\href {\doibase 10.3389/fphy.2021.663457} {\bibfield  {journal}
  {\bibinfo  {journal} {Front. Phys.}\ }\textbf {\bibinfo {volume} {9}},\
  \bibinfo {pages} {663457} (\bibinfo {year} {2021})}\BibitemShut {NoStop}%
\bibitem [{\citenamefont {Engel}\ \emph {et~al.}(2013)\citenamefont {Engel},
  \citenamefont {Anderson}, \citenamefont {Glotzer}, \citenamefont {Isobe},
  \citenamefont {Bernard},\ and\ \citenamefont {Krauth}}]{Engel13}%
  \BibitemOpen
  \bibfield  {author} {\bibinfo {author} {\bibfnamefont {M.}~\bibnamefont
  {Engel}}, \bibinfo {author} {\bibfnamefont {J.~A.}\ \bibnamefont {Anderson}},
  \bibinfo {author} {\bibfnamefont {S.~C.}\ \bibnamefont {Glotzer}}, \bibinfo
  {author} {\bibfnamefont {M.}~\bibnamefont {Isobe}}, \bibinfo {author}
  {\bibfnamefont {E.~P.}\ \bibnamefont {Bernard}}, \ and\ \bibinfo {author}
  {\bibfnamefont {W.}~\bibnamefont {Krauth}},\ }\href {\doibase
  10.1103/PhysRevE.87.042134} {\bibfield  {journal} {\bibinfo  {journal} {Phys.
  Rev. E}\ }\textbf {\bibinfo {volume} {87}},\ \bibinfo {pages} {042134}
  (\bibinfo {year} {2013})}\BibitemShut {NoStop}%
\bibitem [{\citenamefont {Isobe}\ and\ \citenamefont {Krauth}(2015)}]{Isobe15}%
  \BibitemOpen
  \bibfield  {author} {\bibinfo {author} {\bibfnamefont {M.}~\bibnamefont
  {Isobe}}\ and\ \bibinfo {author} {\bibfnamefont {W.}~\bibnamefont {Krauth}},\
  }\href {\doibase 10.1063/1.4929529} {\bibfield  {journal} {\bibinfo
  {journal} {J. Chem. Phys.}\ }\textbf {\bibinfo {volume} {143}},\ \bibinfo
  {pages} {084509} (\bibinfo {year} {2015})}\BibitemShut {NoStop}%
\bibitem [{\citenamefont {Harland}\ \emph {et~al.}(2017)\citenamefont
  {Harland}, \citenamefont {Michel}, \citenamefont {Kampmann},\ and\
  \citenamefont {Kierfeld}}]{Harland17}%
  \BibitemOpen
  \bibfield  {author} {\bibinfo {author} {\bibfnamefont {J.}~\bibnamefont
  {Harland}}, \bibinfo {author} {\bibfnamefont {M.}~\bibnamefont {Michel}},
  \bibinfo {author} {\bibfnamefont {T.~A.}\ \bibnamefont {Kampmann}}, \ and\
  \bibinfo {author} {\bibfnamefont {J.}~\bibnamefont {Kierfeld}},\ }\href
  {\doibase 10.1209/0295-5075/117/30001} {\bibfield  {journal} {\bibinfo
  {journal} {Europhys. Lett.}\ }\textbf {\bibinfo {volume} {117}},\ \bibinfo
  {pages} {30001} (\bibinfo {year} {2017})}\BibitemShut {NoStop}%
\bibitem [{\citenamefont {Kapfer}\ and\ \citenamefont
  {Krauth}(2016)}]{Kapfer16}%
  \BibitemOpen
  \bibfield  {author} {\bibinfo {author} {\bibfnamefont {S.~C.}\ \bibnamefont
  {Kapfer}}\ and\ \bibinfo {author} {\bibfnamefont {W.}~\bibnamefont
  {Krauth}},\ }\href {\doibase 10.1103/PhysRevE.94.031302} {\bibfield
  {journal} {\bibinfo  {journal} {Phys. Rev. E}\ }\textbf {\bibinfo {volume}
  {94}},\ \bibinfo {pages} {031302} (\bibinfo {year} {2016})}\BibitemShut
  {NoStop}%
\bibitem [{\citenamefont {Michel}, \citenamefont {Mayer},\ and\ \citenamefont
  {Krauth}(2015)}]{Michel2015}%
  \BibitemOpen
  \bibfield  {author} {\bibinfo {author} {\bibfnamefont {M.}~\bibnamefont
  {Michel}}, \bibinfo {author} {\bibfnamefont {J.}~\bibnamefont {Mayer}}, \
  and\ \bibinfo {author} {\bibfnamefont {W.}~\bibnamefont {Krauth}},\ }\href
  {\doibase 10.1209/0295-5075/112/20003} {\bibfield  {journal} {\bibinfo
  {journal} {Europhy. Lett.}\ }\textbf {\bibinfo {volume} {112}},\ \bibinfo
  {pages} {20003} (\bibinfo {year} {2015})}\BibitemShut {NoStop}%
\bibitem [{\citenamefont {Prokof’ev}, \citenamefont {Svistunov},\ and\
  \citenamefont {Tupitsyn}(1998)}]{prokofev_exact_1998}%
  \BibitemOpen
  \bibfield  {author} {\bibinfo {author} {\bibfnamefont {N.~V.}\ \bibnamefont
  {Prokof’ev}}, \bibinfo {author} {\bibfnamefont {B.~V.}\ \bibnamefont
  {Svistunov}}, \ and\ \bibinfo {author} {\bibfnamefont {I.~S.}\ \bibnamefont
  {Tupitsyn}},\ }\href {\doibase 10.1134/1.558661} {\bibfield  {journal}
  {\bibinfo  {journal} {J. Exp. Theor. Phys.}\ }\textbf {\bibinfo {volume}
  {87}},\ \bibinfo {pages} {310} (\bibinfo {year} {1998})}\BibitemShut
  {NoStop}%
\bibitem [{\citenamefont {Boninsegni}, \citenamefont {Prokof’ev},\ and\
  \citenamefont {Svistunov}(2006)}]{boninsegni_worm_2006}%
  \BibitemOpen
  \bibfield  {author} {\bibinfo {author} {\bibfnamefont {M.}~\bibnamefont
  {Boninsegni}}, \bibinfo {author} {\bibfnamefont {N.~V.}\ \bibnamefont
  {Prokof’ev}}, \ and\ \bibinfo {author} {\bibfnamefont {B.~V.}\ \bibnamefont
  {Svistunov}},\ }\href {\doibase 10.1103/PhysRevE.74.036701} {\bibfield
  {journal} {\bibinfo  {journal} {Phys. Rev. E}\ }\textbf {\bibinfo {volume}
  {74}},\ \bibinfo {pages} {036701} (\bibinfo {year} {2006})}\BibitemShut
  {NoStop}%
\bibitem [{\citenamefont {Syljuåsen}\ and\ \citenamefont
  {Sandvik}(2002)}]{syljuasen_quantum_2002}%
  \BibitemOpen
  \bibfield  {author} {\bibinfo {author} {\bibfnamefont {O.~F.}\ \bibnamefont
  {Syljuåsen}}\ and\ \bibinfo {author} {\bibfnamefont {A.~W.}\ \bibnamefont
  {Sandvik}},\ }\href {\doibase 10.1103/PhysRevE.66.046701} {\bibfield
  {journal} {\bibinfo  {journal} {Phys. Rev. E}\ }\textbf {\bibinfo {volume}
  {66}},\ \bibinfo {pages} {046701} (\bibinfo {year} {2002})}\BibitemShut
  {NoStop}%
\bibitem [{\citenamefont {Prokof'ev}\ and\ \citenamefont
  {Svistunov}(2001)}]{prokofev_worm_2001}%
  \BibitemOpen
  \bibfield  {author} {\bibinfo {author} {\bibfnamefont {N.}~\bibnamefont
  {Prokof'ev}}\ and\ \bibinfo {author} {\bibfnamefont {B.}~\bibnamefont
  {Svistunov}},\ }\href {\doibase 10.1103/PhysRevLett.87.160601} {\bibfield
  {journal} {\bibinfo  {journal} {Phys. Rev. Lett.}\ }\textbf {\bibinfo
  {volume} {87}},\ \bibinfo {pages} {160601} (\bibinfo {year}
  {2001})}\BibitemShut {NoStop}%
\bibitem [{\citenamefont {Wang}(2005)}]{wang_worm_2005}%
  \BibitemOpen
  \bibfield  {author} {\bibinfo {author} {\bibfnamefont {J.-S.}\ \bibnamefont
  {Wang}},\ }\href {\doibase 10.1103/PhysRevE.72.036706} {\bibfield  {journal}
  {\bibinfo  {journal} {Phys. Rev. E}\ }\textbf {\bibinfo {volume} {72}},\
  \bibinfo {pages} {036706} (\bibinfo {year} {2005})}\BibitemShut {NoStop}%
\bibitem [{\citenamefont {Liu}, \citenamefont {Deng},\ and\ \citenamefont
  {Garoni}(2011)}]{liu_worm_2011}%
  \BibitemOpen
  \bibfield  {author} {\bibinfo {author} {\bibfnamefont {Q.}~\bibnamefont
  {Liu}}, \bibinfo {author} {\bibfnamefont {Y.}~\bibnamefont {Deng}}, \ and\
  \bibinfo {author} {\bibfnamefont {T.~M.}\ \bibnamefont {Garoni}},\ }\href
  {\doibase 10.1016/j.nuclphysb.2011.01.003} {\bibfield  {journal} {\bibinfo
  {journal} {Nucl. Phys. B.}\ }\textbf {\bibinfo {volume} {846}},\ \bibinfo
  {pages} {283} (\bibinfo {year} {2011})}\BibitemShut {NoStop}%
\bibitem [{\citenamefont {Adams}\ and\ \citenamefont
  {Chandrasekharan}(2003)}]{adams_chiral_2003}%
  \BibitemOpen
  \bibfield  {author} {\bibinfo {author} {\bibfnamefont {D.~H.}\ \bibnamefont
  {Adams}}\ and\ \bibinfo {author} {\bibfnamefont {S.}~\bibnamefont
  {Chandrasekharan}},\ }\href {\doibase 10.1016/S0550-3213(03)00350-X}
  {\bibfield  {journal} {\bibinfo  {journal} {Nucl. Phys. B.}\ }\textbf
  {\bibinfo {volume} {662}},\ \bibinfo {pages} {220} (\bibinfo {year}
  {2003})}\BibitemShut {NoStop}%
\bibitem [{\citenamefont {Suwa}(2021)}]{suwa_geometric_2021}%
  \BibitemOpen
  \bibfield  {author} {\bibinfo {author} {\bibfnamefont {H.}~\bibnamefont
  {Suwa}},\ }\href {\doibase 10.1103/PhysRevE.103.013308} {\bibfield  {journal}
  {\bibinfo  {journal} {Phys. Rev. E}\ }\textbf {\bibinfo {volume} {103}},\
  \bibinfo {pages} {013308} (\bibinfo {year} {2021})}\BibitemShut {NoStop}%
\bibitem [{\citenamefont {Suwa}(2022)}]{suwa_lifted_2022}%
  \BibitemOpen
  \bibfield  {author} {\bibinfo {author} {\bibfnamefont {H.}~\bibnamefont
  {Suwa}},\ }\href {\doibase 10.1103/PhysRevE.106.055306} {\bibfield  {journal}
  {\bibinfo  {journal} {Phys. Rev. E}\ }\textbf {\bibinfo {volume} {106}},\
  \bibinfo {pages} {055306} (\bibinfo {year} {2022})}\BibitemShut {NoStop}%
\bibitem [{\citenamefont {van~der Waerden}(1941)}]{Waerden1941}%
  \BibitemOpen
  \bibfield  {author} {\bibinfo {author} {\bibfnamefont {B.~L.}\ \bibnamefont
  {van~der Waerden}},\ }\href {\doibase 10.1007/BF01342928} {\bibfield
  {journal} {\bibinfo  {journal} {Zeitschrift f{\"u}r Physik}\ }\textbf
  {\bibinfo {volume} {118}},\ \bibinfo {pages} {473} (\bibinfo {year}
  {1941})}\BibitemShut {NoStop}%
\bibitem [{\citenamefont {Wolff}(1989)}]{wolff_collective_1989}%
  \BibitemOpen
  \bibfield  {author} {\bibinfo {author} {\bibfnamefont {U.}~\bibnamefont
  {Wolff}},\ }\href {\doibase 10.1103/PhysRevLett.62.361} {\bibfield  {journal}
  {\bibinfo  {journal} {Phys. Rev. Lett.}\ }\textbf {\bibinfo {volume} {62}},\
  \bibinfo {pages} {361} (\bibinfo {year} {1989})}\BibitemShut {NoStop}%
\bibitem [{\citenamefont {Liu}, \citenamefont {Polkovnikov},\ and\
  \citenamefont {Sandvik}(2014)}]{Liu2014}%
  \BibitemOpen
  \bibfield  {author} {\bibinfo {author} {\bibfnamefont {C.-W.}\ \bibnamefont
  {Liu}}, \bibinfo {author} {\bibfnamefont {A.}~\bibnamefont {Polkovnikov}}, \
  and\ \bibinfo {author} {\bibfnamefont {A.~W.}\ \bibnamefont {Sandvik}},\
  }\href {\doibase 10.1103/PhysRevB.89.054307} {\bibfield  {journal} {\bibinfo
  {journal} {Phys. Rev. B}\ }\textbf {\bibinfo {volume} {89}},\ \bibinfo
  {pages} {054307} (\bibinfo {year} {2014})}\BibitemShut {NoStop}%
\bibitem [{\citenamefont {Li}\ \emph {et~al.}(2021)\citenamefont {Li},
  \citenamefont {Todo}, \citenamefont {Maggs},\ and\ \citenamefont
  {Krauth}}]{li_multithreaded_2021}%
  \BibitemOpen
  \bibfield  {author} {\bibinfo {author} {\bibfnamefont {B.}~\bibnamefont
  {Li}}, \bibinfo {author} {\bibfnamefont {S.}~\bibnamefont {Todo}}, \bibinfo
  {author} {\bibfnamefont {A.~C.}\ \bibnamefont {Maggs}}, \ and\ \bibinfo
  {author} {\bibfnamefont {W.}~\bibnamefont {Krauth}},\ }\href {\doibase
  10.1016/j.cpc.2020.107702} {\bibfield  {journal} {\bibinfo  {journal}
  {Comput. Phys. Commun.}\ }\textbf {\bibinfo {volume} {261}},\ \bibinfo
  {pages} {107702} (\bibinfo {year} {2021})}\BibitemShut {NoStop}%
\bibitem [{\citenamefont {Elçi}\ \emph {et~al.}(2018)\citenamefont {Elçi},
  \citenamefont {Grimm}, \citenamefont {Ding}, \citenamefont {Nasrawi},
  \citenamefont {Garoni},\ and\ \citenamefont {Deng}}]{elci_lifted_2018}%
  \BibitemOpen
  \bibfield  {author} {\bibinfo {author} {\bibfnamefont {E.~M.}\ \bibnamefont
  {Elçi}}, \bibinfo {author} {\bibfnamefont {J.}~\bibnamefont {Grimm}},
  \bibinfo {author} {\bibfnamefont {L.}~\bibnamefont {Ding}}, \bibinfo {author}
  {\bibfnamefont {A.}~\bibnamefont {Nasrawi}}, \bibinfo {author} {\bibfnamefont
  {T.~M.}\ \bibnamefont {Garoni}}, \ and\ \bibinfo {author} {\bibfnamefont
  {Y.}~\bibnamefont {Deng}},\ }\href {\doibase 10.1103/PhysRevE.97.042126}
  {\bibfield  {journal} {\bibinfo  {journal} {Phys. Rev. E}\ }\textbf {\bibinfo
  {volume} {97}},\ \bibinfo {pages} {042126} (\bibinfo {year}
  {2018})}\BibitemShut {NoStop}%
\bibitem [{\citenamefont {Sakai}\ and\ \citenamefont
  {Hukushima}(2016)}]{sakai_irreversible_2016}%
  \BibitemOpen
  \bibfield  {author} {\bibinfo {author} {\bibfnamefont {Y.}~\bibnamefont
  {Sakai}}\ and\ \bibinfo {author} {\bibfnamefont {K.}~\bibnamefont
  {Hukushima}},\ }\href {\doibase 10.7566/JPSJ.85.104002} {\bibfield  {journal}
  {\bibinfo  {journal} {J. Phys. Soc. Jpn.}\ }\textbf {\bibinfo {volume}
  {85}},\ \bibinfo {pages} {104002} (\bibinfo {year} {2016})}\BibitemShut
  {NoStop}%
\bibitem [{\citenamefont {Syed}\ \emph {et~al.}(2022)\citenamefont {Syed},
  \citenamefont {Bouchard-Côté}, \citenamefont {Deligiannidis},\ and\
  \citenamefont {Doucet}}]{syed_non-reversible_2022}%
  \BibitemOpen
  \bibfield  {author} {\bibinfo {author} {\bibfnamefont {S.}~\bibnamefont
  {Syed}}, \bibinfo {author} {\bibfnamefont {A.}~\bibnamefont
  {Bouchard-Côté}}, \bibinfo {author} {\bibfnamefont {G.}~\bibnamefont
  {Deligiannidis}}, \ and\ \bibinfo {author} {\bibfnamefont {A.}~\bibnamefont
  {Doucet}},\ }\href {\doibase 10.1111/rssb.12464} {\bibfield  {journal}
  {\bibinfo  {journal} {J. R. Stat. Soc. Ser. B Stat. Method.}\ }\textbf
  {\bibinfo {volume} {84}},\ \bibinfo {pages} {321} (\bibinfo {year}
  {2022})}\BibitemShut {NoStop}%
\bibitem [{\citenamefont {Itoh}\ and\ \citenamefont
  {Okumura}(2013)}]{itoh_replica-permutation_2013}%
  \BibitemOpen
  \bibfield  {author} {\bibinfo {author} {\bibfnamefont {S.~G.}\ \bibnamefont
  {Itoh}}\ and\ \bibinfo {author} {\bibfnamefont {H.}~\bibnamefont {Okumura}},\
  }\href {\doibase 10.1021/ct3007919} {\bibfield  {journal} {\bibinfo
  {journal} {J. Chem. Theory Comput.}\ }\textbf {\bibinfo {volume} {9}},\
  \bibinfo {pages} {570} (\bibinfo {year} {2013})}\BibitemShut {NoStop}%
\bibitem [{\citenamefont {Bierkens}(2016)}]{bierkens_non-reversible_2016}%
  \BibitemOpen
  \bibfield  {author} {\bibinfo {author} {\bibfnamefont {J.}~\bibnamefont
  {Bierkens}},\ }\href {\doibase 10.1007/s11222-015-9598-x} {\bibfield
  {journal} {\bibinfo  {journal} {Stat. Comput.}\ }\textbf {\bibinfo {volume}
  {26}},\ \bibinfo {pages} {1213} (\bibinfo {year} {2016})}\BibitemShut
  {NoStop}%
\bibitem [{\citenamefont {Ohzeki}\ and\ \citenamefont
  {Ichiki}(2015)}]{ohzeki_langevin_2015}%
  \BibitemOpen
  \bibfield  {author} {\bibinfo {author} {\bibfnamefont {M.}~\bibnamefont
  {Ohzeki}}\ and\ \bibinfo {author} {\bibfnamefont {A.}~\bibnamefont
  {Ichiki}},\ }\href {\doibase 10.1103/PhysRevE.92.012105} {\bibfield
  {journal} {\bibinfo  {journal} {Phys. Rev. E}\ }\textbf {\bibinfo {volume}
  {92}},\ \bibinfo {pages} {012105} (\bibinfo {year} {2015})}\BibitemShut
  {NoStop}%
\bibitem [{\citenamefont {Coghi}, \citenamefont {Chetrite},\ and\ \citenamefont
  {Touchette}(2021)}]{coghi_role_2021}%
  \BibitemOpen
  \bibfield  {author} {\bibinfo {author} {\bibfnamefont {F.}~\bibnamefont
  {Coghi}}, \bibinfo {author} {\bibfnamefont {R.}~\bibnamefont {Chetrite}}, \
  and\ \bibinfo {author} {\bibfnamefont {H.}~\bibnamefont {Touchette}},\ }\href
  {\doibase 10.1103/PhysRevE.103.062142} {\bibfield  {journal} {\bibinfo
  {journal} {Phys. Rev. E}\ }\textbf {\bibinfo {volume} {103}},\ \bibinfo
  {pages} {062142} (\bibinfo {year} {2021})}\BibitemShut {NoStop}%
\bibitem [{\citenamefont {Abdulle}, \citenamefont {Pavliotis},\ and\
  \citenamefont {Vilmart}(2019)}]{abdulle_accelerated_2019}%
  \BibitemOpen
  \bibfield  {author} {\bibinfo {author} {\bibfnamefont {A.}~\bibnamefont
  {Abdulle}}, \bibinfo {author} {\bibfnamefont {G.~A.}\ \bibnamefont
  {Pavliotis}}, \ and\ \bibinfo {author} {\bibfnamefont {G.}~\bibnamefont
  {Vilmart}},\ }\href {\doibase 10.1016/j.crma.2019.04.008} {\bibfield
  {journal} {\bibinfo  {journal} {Comptes Rendus. Mathématique}\ }\textbf
  {\bibinfo {volume} {357}},\ \bibinfo {pages} {349} (\bibinfo {year}
  {2019})}\BibitemShut {NoStop}%
\bibitem [{\citenamefont {Wang}\ \emph {et~al.}(2021)\citenamefont {Wang},
  \citenamefont {Pollock}, \citenamefont {Roberts},\ and\ \citenamefont
  {Steinsaltz}}]{wang_regeneration-enriched_2021}%
  \BibitemOpen
  \bibfield  {author} {\bibinfo {author} {\bibfnamefont {A.~Q.}\ \bibnamefont
  {Wang}}, \bibinfo {author} {\bibfnamefont {M.}~\bibnamefont {Pollock}},
  \bibinfo {author} {\bibfnamefont {G.~O.}\ \bibnamefont {Roberts}}, \ and\
  \bibinfo {author} {\bibfnamefont {D.}~\bibnamefont {Steinsaltz}},\ }\href
  {\doibase 10.1214/20-AAP1602} {\bibfield  {journal} {\bibinfo  {journal}
  {Ann. Appl. Probab.}\ }\textbf {\bibinfo {volume} {31}},\ \bibinfo {pages}
  {703} (\bibinfo {year} {2021})}\BibitemShut {NoStop}%
\end{thebibliography}%

\end{document}